\pdfoutput=1
\documentclass{aa}

\usepackage{xspace} 
\usepackage{aas_macros} 
\usepackage[load-configurations = abbreviations,load=accepted]{siunitx} 
\usepackage{lscape}
\usepackage{longtable}
\usepackage{tabularx}

\usepackage{multirow}

\usepackage{subfig,graphicx}
\usepackage{txfonts}
\usepackage[draft,colorlinks=true, citecolor={blue}]{hyperref}

\newcommand{\wise}{WISE\xspace}
\newcommand{\fermi}{Fermi\xspace}
\newcommand{\lat}{LAT\xspace}
\newcommand{\bll}{BL\,Lac\xspace}
\newcommand{\blls}{BL\,Lacs\xspace}
\newcommand{\fsrq}{FSRQ\xspace}
\newcommand{\fsrqs}{FSRQs\xspace}
\newcommand{\tmva}{TMVA\xspace}
\newcommand{\supernovae}{supernov\ae\xspace}



\begin{document} 

   \title{Research and characterisation of blazar candidates among the
     Fermi/LAT 3FGL catalogue using multivariate classifications}

   \author{Julien Lefaucheur\inst{1,2} \and Santiago Pita\inst{2}}

   \institute{LUTH, Observatoire de Paris, PSL Research University, CNRS, Universit\'e Paris Diderot,
     5 Place Jules Janssen, 92190 Meudon, France \label{inst2} \and
     APC, AstroParticule et Cosmologie, Universit\'{e} Paris Diderot, CNRS/IN2P3, CEA/Irfu,
     Observatoire de Paris, Sorbonne Paris Cit\'{e}, 10,
     rue Alice Domon et L\'{e}onie Duquet, 75205 Paris Cedex 13, France\label{inst2} \\
     \email{julien.lefaucheur@obspm.fr or santiago.pita@apc.in2p3.fr}\label{inst2}}

   \date{Received 19/08/2016; accepted 20/02/2017}
 
   \abstract
       {In the recently published 3FGL catalogue, the Fermi/LAT collaboration
         reports the detection of $\gamma$-ray emission from 3034 sources
         obtained after four years of observations.
         The nature of 1010 of those sources is unknown, 
         whereas 2023 have well-identified counterparts in other wavelengths.
         Most of the associated sources are labelled as blazars (1717/2023),
         but the \bll or \fsrq 
         nature of 573 of these
         blazars is still undetermined.} 
       {The aim of this study was two-fold. First, to significantly increase
         the number of blazar candidates from a search among the large number of
         Fermi/LAT 3FGL unassociated sources (case A).
         Second, to determine the \bll or \fsrq nature of the blazar candidates,
         including those determined as such in this work and the blazar candidates of 
         uncertain type (BCU) that are already present in the 3FGL catalogue (case B).}
       {For this purpose, multivariate classifiers -- boosted decision trees
         and multilayer perceptron neural networks -- were trained using samples of labelled sources with no caution flag from the 3FGL catalogue and
         carefully chosen discriminant parameters.
         The decisions of the classifiers were combined in order to obtain a high
         level of source identification along with well controlled numbers of 
         expected false associations.
         Specifically for case A, dedicated classifications were generated for
         high ($|b|>\SI{10}{\degree}$) and low  ($|b| \leq \SI{10}{\degree}$) galactic latitude sources; in addition, the application of classifiers to samples of sources with caution flag was considered separately, and specific performance metrics were estimated.
       }
       {We obtained a sample of 595 blazar candidates (high and low galactic latitude) among
         the unassociated sources of the 3FGL catalogue. 
         We also obtained a sample of 509 \blls and 295 \fsrqs from the blazar candidates cited above and the 
         BCUs of the 3FGL catalogue.
         The number of expected false associations
         is given for different samples of candidates.  
         It is, in particular, notably low ($\sim$9/425) for the sample of high-latitude blazar candidates from case A.}
       {}


  \titlerunning{Research and characterisation of blazar candidates among the Fermi/LAT 3FGL catalogue}
  \maketitle
%

\section{Introduction}

The LAT telescope, on board the Fermi satellite, has been mapping the $\gamma$-ray sky
(above 100 MeV) since 2008 with unprecedented angular resolution and sensitivity.
In the recently published 3FGL catalogue \citep{Acero:2015aa},
the \fermi/\lat collaboration reports the detection of $\gamma$-ray 
emission from \SI{3034}{} sources above $4\sigma$ significance,
obtained after four years of observations.
Among these sources, 2025 have been associated\footnote{Including the so-called
  identified sources showing periodic emission, variability, or extended morphology
  in different wavelengths.}
with sources of well known types detected at other wavelengths.
Most of them are active galactic nuclei (AGN) (1752), and, of particular interest here,
blazars (1717), among which 660 are labelled as BL Lacertae objects (\bll), 484 as
flat spectrum radio quasar (FSRQ)
and 573 as blazars of undetermined type.
The remaining fraction is composed of galactic sources, mainly pulsars (166) and also
\supernovae remnants or pulsar wind nebul\ae\xspace (85).
Nevertheless, one third of the 3FGL catalogue sources are still of unknown nature 
because of the lack of firmly identified counterparts at other wavelengths.
It is likely that a significant fraction of these unassociated sources are blazars, considering the incompleteness of counterpart catalogues,
  the existence of $\gamma$-ray sources with multiple candidate associations due to the large error localisation of the \fermi/LAT,
  and also a deficit seen at low values ($|b| \leq \SI{10}{\degree}$) in the latitude distribution of Fermi blazars \citep{Acero:2015aa}.

The understanding of the blazar population and its evolution --~for example the
validity of the ``blazar sequence''~-- and the determination of the extragalactic
background light (EBL) are key topics in high-energy astrophysics \citep{Sol:2013aa}
which are currently limited, observationally, by the small number of
detected blazars.
For this reason, several studies have addressed the question of the nature of the Fermi/LAT catalogues'
sources of unknown type. Two different approaches are generally used, one based on machine-learning classification methods
and the other based on multiwavelength identifications or associations, they are described below.

The first approach is based on the exploitation of statistical
differences imprinted in the Fermi/LAT catalogues, 
such as variability and spectral shape, between different populations of sources.
Classifications are built with machine-learning algorithms, using given sets of
discriminant parameters, to search for particular types of sources among the unassociated ones.
\citet{Ackermann:2012ab} identified AGN and pulsar candidates among the
630 unassociated sources of the 1FGL catalogue \citep{Abdo:2010ab} with a
classification built on the decisions of two individual classifiers based
on random forest and logical regression multivariate methods.
They proposed a list of 221 AGN and 134 pulsar candidates.
To search for possible dark matter candidates in the sample of the 269 unassociated sources
located at high galactic latitude ($|b| > \SI{10}{\degree}$) of the 2FGL catalogue \citep{Nolan:2012aa},
\citet{Mirabal2012} focused on the outliers of their own AGN and pulsar classifications
built with a random forest method.
They proposed a list of 216 AGN candidates.
\citet{Hassan:2013aa} then identified 235 possible \bll or \fsrq candidates among
the 269 blazars of unknown type in the 2FGL catalogue by combining
the decisions of two classifiers based on support vector machine
and random forest methods.
In another study, using a combination of neural network and random forest methods,
and introducing new strongly-discriminant parameters, 
\citet{Doert:2013aa} identified a sample of 231 AGN candidates among 576
unassociated sources of the 2FGL catalogue.
Recently, \citet{Saz-Parkinson:2016aa} applied a random forest and a logistic regression
algorithm to identify pulsar and AGN candidates
among the unassociated sources in the 3FGL catalogue.
They proposed a list of 334 pulsar candidates and 559 AGN candidates.
Finally, \citet{Chiaro:2016aa} applied a neural network to identify \blls
  and \fsrqs among the blazar candidates of uncertain type (BCU) in the 3FGL
  catalogue.
  They obtained a list of 314 \bll candidates and 113 \fsrq candidates.

The second approach consists of finding possible counterparts in
different wavelength bands,
beyond what was done by the Fermi/LAT collaboration for their public catalogues.
Aside from determining the nature of the source, the better localisation
of candidate counterparts simplifies more detailed identification
efforts at other wavelengths.
A first attempt by \citet{Massaro:2011aa}
used the assumption that blazars occupy a special position in
the colour-colour diagram constructed with the first three
filters of the \wise satellite \citep{Wright:2010aa}.
By building ``blazar'' regions with a selected sample of infrared blazars,
and by comparing the distance of the unassociated sources in the colour
space to these regions, one can identify candidates for blazar-like counterparts.
This method has been improved
several times and applied to the 2FGL catalogue
\citep{Massaro:2012ac,Massaro:2012aa,Massaro:2013ab}, thus providing lists
of possible blazar counterparts.
In \citet{Massaro:2013ab} the authors provide
149 infrared counterparts corresponding to 109 2FGL unassociated sources.
There is, however, no estimate of the number of false associations,
as the method is based only on a selected sample of blazars and does not consider
the behaviour of other infrared source classes.
Source contamination in searches for counterparts in infrared catalogues
is illustrated in \citet{DAbrusco:2014aa}.
Other attempts have been made with non-parametric techniques, such as kernel
density estimators, using additional information obtained in radio \citep{Massaro:2013aa,Massaro:2014aa} 
or $X$-rays \citep{Paggi:2014aa},
to identify potential blazar counterparts 
for a few tens of unassociated sources in the 2FGL catalogue.
Finally, one can deal with unassociated sources individually,
this is done in the study of \citet{Acero:2013aa} for a limited sample of sources,
by combining multiwavelength observations and analysing the spectral energy
distributions of the sources.

The aim of this study is two-fold. 
First, to significantly increase the number of $\gamma$-ray blazar candidates 
from a search among the large number of \fermi/\lat 3FGL unassociated sources (case A).
Second, to determine the nature (\bll or \fsrq) of the blazar candidates, 
including those determined as such in this work and those labelled as BCU (blazar candidates 
of uncertain type) in the 3FGL catalogue (case B).
For each case, classifiers based on two different machine-learning algorithms were built using only parameters 
from the 3FGL catalogue and combined in order to increase the overall performance. 
Specifically for case A, those classifiers were trained separately for
low ($|b| \leq \SI{10}{\degree}$) and high ($|b| > \SI{10}{\degree}$) galactic latitudes.
Special attention was devoted to the estimation of their performance metrics.
This paper is organised as follows.
In Section \ref{Sec:FermiFermiSamples} we describe the samples of sources used in this work, and also
the selected sets of discriminant parameters.
In Section \ref{Sec:FermiFermiMVA} we present two selected machine-learning algorithms and their settings.
In Section \ref{Sec:TrainingAndPerf} we describe the training of the classifiers and performance evaluation.
Results are then presented in Section \ref{Sec:FermiFermiResults} and discussed in Section \ref{Sec:Discussion}.

\section{Data samples and discriminant parameters}
\label{Sec:FermiFermiSamples}

\subsection{Data samples for classifier building}
\label{Sec:DataSamples}

\begin{table*}
  \caption{Number of sources used to build each classifier and to derive its
    performance metrics.
    The number of sources to which each classifier was applied
    is also given: these ``targets'' correspond to unassociated sources (UnIds) for case A or blazars of unknown type (BCUs) for case B.
    HL and LL refer to the high and low galactic latitude studies (case A), respectively.}              
\label{tab:Pop}      
\centering                                      
\begin{tabular}{l c c c c c c }          

  \cline{1-7}
  \multicolumn{1}{|c}{} & \multicolumn{3}{|c|}{Samples with no flag} & \multicolumn{3}{c|}{Samples with flags} \\      
  \hline
  \hline
  \multicolumn{1}{|c}{Studies} & \multicolumn{2}{|c}{Training (70\%) and Test (30\%)} & \multicolumn{1}{|c|}{Targets} & \multicolumn{2}{c|}{Test} & \multicolumn{1}{c|}{Targets}  \\      

  \hline
  \hline                                   
  \multicolumn{1}{|l}{case~A (HL)} &
  \multicolumn{1}{|l}{blazars (1572)} &  \multicolumn{1}{|l|}{pulsars (134)} & \multicolumn{1}{l|}{ UnIds (422)} &
  \multicolumn{1}{l}{blazars (145)} &  \multicolumn{1}{|l|}{pulsars (32)} & \multicolumn{1}{l|}{UnIds (109)} \\

  \hline                                   
  \multicolumn{1}{|l}{case~A (LL)} &
  \multicolumn{1}{|l}{blazars (1572) } &  \multicolumn{1}{|l|}{galactic sources (183)} & \multicolumn{1}{l|}{UnIds (169)} &
  \multicolumn{1}{l}{blazars (145)} &  \multicolumn{1}{|l|}{galactic sources (75)}  & \multicolumn{1}{l|}{ UnIds (247)} \\

  \hline                                   
  \multicolumn{1}{|l}{case B} &
  \multicolumn{1}{|l}{ \blls (638)} &  \multicolumn{1}{|l|}{\fsrqs (448)} & \multicolumn{1}{l|}{BCUs (486)} &
  \multicolumn{1}{c}{--} &  \multicolumn{1}{|c|}{--} & \multicolumn{1}{c|}{--} \\

  \hline
  
\end{tabular}
\tablefoot{Only BCUs are counted in the target sample for case B, the corresponding classifiers will be also applied to blazar candidates from case A.}
\end{table*}

The aim of the first study (case A) is to identify blazar candidates
among the unassociated sources of the 3FGL catalogue \citep{Acero:2015aa}.
Considering that at high galactic latitudes the unassociated sources are likely to be either blazars or pulsars \citep{Mirabal2012},
classifiers were built and tested using a sample of \SI{1572}{} blazars (including \blls, \fsrqs and BCUs)
and a sample of \SI{134}{} pulsars, regardless of their galactic latitudes.
On the other hand, as at low galactic latitudes the unassociated sources are likely to be blazars or any type of galactic sources,
other classifiers were built and tested using the same sample of \SI{1572}{} blazars and a sample of \SI{183}{} galactic sources,
corresponding to \SI{134}{} pulsars, \SI{34}{} pulsar wind nebulae (PWN) or supernova remnants (SNR), and also a few globular clusters and binaries.
Only sources that have no caution flags\footnote{The flags in the 3FGL catalogue indicate that a possible problem arose during
  the analysis of the $\gamma$-ray sources \citep{Acero:2015aa}.} in the 3FGL catalogue were considered.
For each case, these samples were split into training and test samples (respectively \SI{70}{\percent}
  and \SI{30}{\percent}) following a procedure explained in Section~\ref{Sec:SplitSamples}.
The test sample was used to determine the performance of the classifiers built with the training sample.
  In addition, a sample of identified or associated flagged sources\footnote{Except the so called ``c-sources''
    which were discarded from the study as they are considered to be potentially confused with galactic diffuse emission.}
  was used only to estimate the performance of the classifiers specifically for flagged sources.
The "high latitude" and "low latitude" classifiers were applied to unassociated sources with galactic latitudes $|b| > \SI{10}{\degree}$
and $|b| \leq \SI{10}{\degree}$, respectively.
Numbers are summarised in Table~\ref{tab:Pop}. 

The aim of the second study (case B) is to determine the nature (\bll or \fsrq)
of blazar candidates in the 3FGL catalogue for which this information is not known.
In this case classifiers were built and tested using a
sample of \SI{638}{} sources labelled as \blls and a sample of \SI{448}{}
sources labelled as \fsrqs in the 3FGL catalogue. Here also, only sources with no flag were considered to build and test the classifiers.
As the flagged sources were few in number (\SI{22}{} and \SI{36}{}, for \blls and \fsrqs respectively),
it was not possible to derive a reliable estimation of their performance when applied to flagged sources.
For this reason, classifiers were applied only to the sample of blazar candidates
of unknown type and with no flag (486 BCUs from the 3FGL catalogue and also the blazar candidates resulting from the case A study).
Numbers are summarised in Table~\ref{tab:Pop}.   

\subsection{Discriminant parameters}
\label{Sec:DiscriParams}

\begin{figure*}
  \centering
  \subfloat[]{
    \label{fig:NormVarVsLambda}
    \includegraphics[width=0.75\columnwidth]{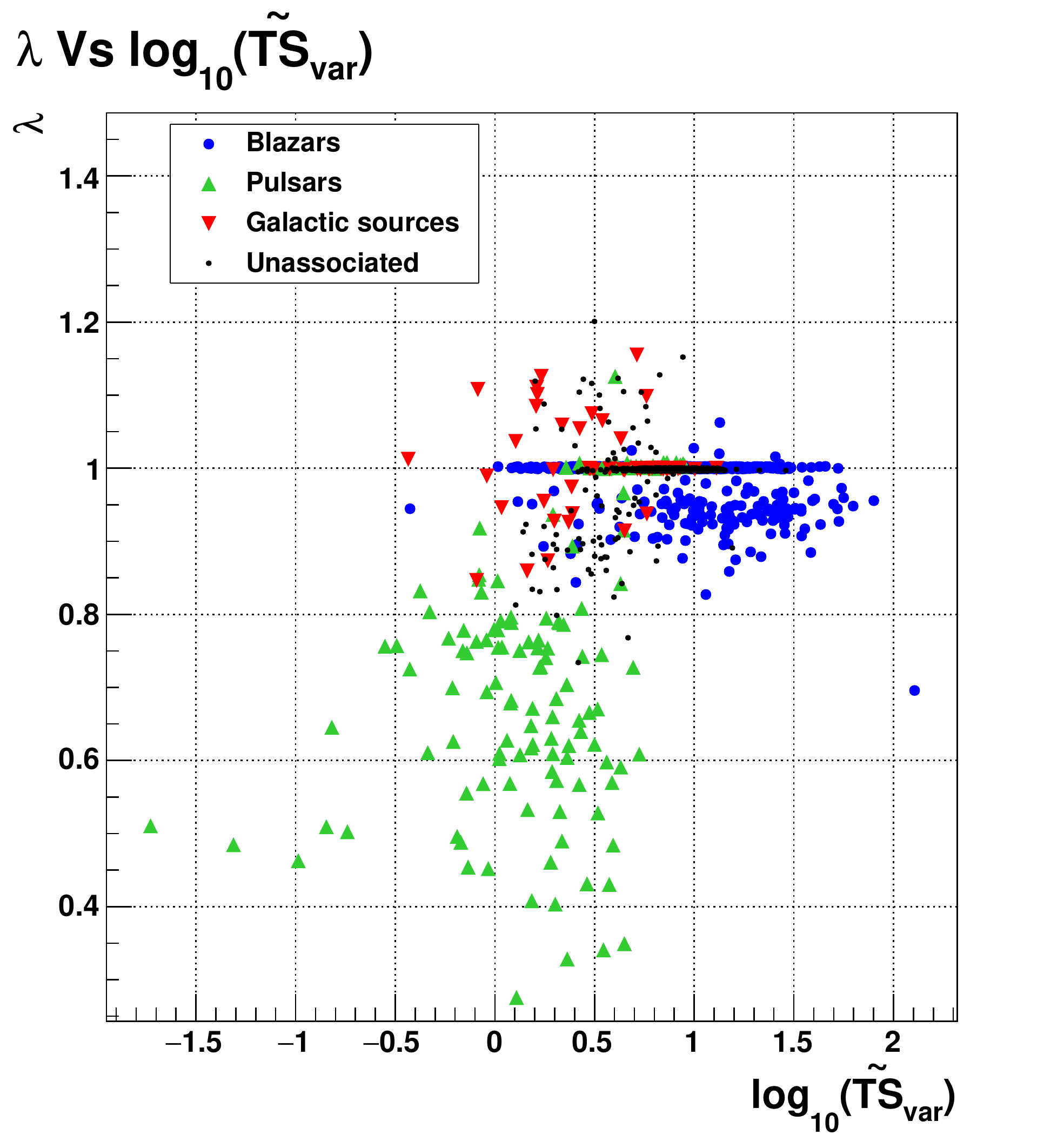}
  }
  \hspace{0.1\linewidth}
  \subfloat[]{
    \label{fig:NormCurvVsNormVar}
    \includegraphics[width=0.75\columnwidth]{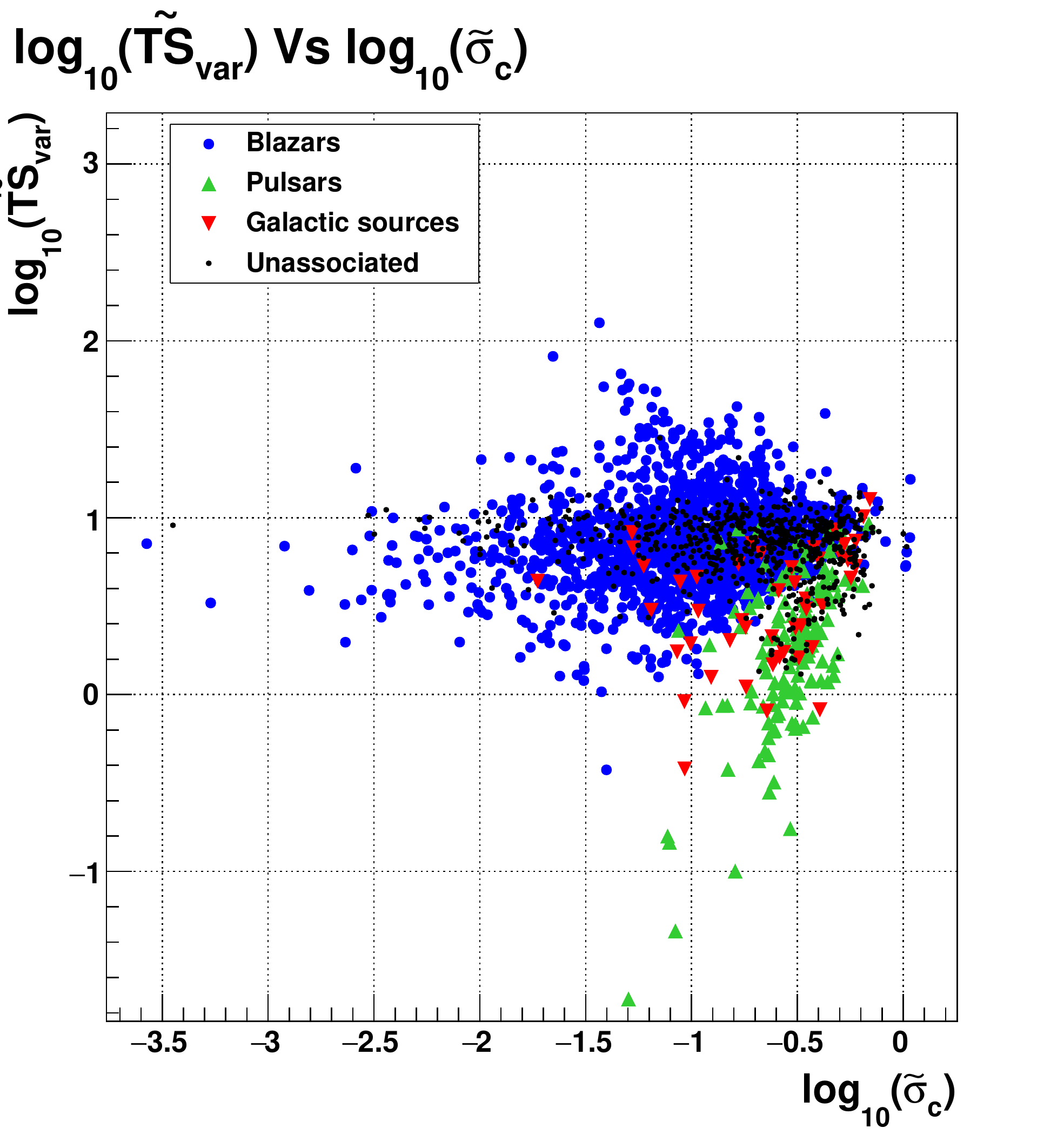}
  }\\[-2ex]

  \subfloat[]{
    \label{fig:LambdaVsNormCurv}
    \includegraphics[width=0.75\columnwidth]{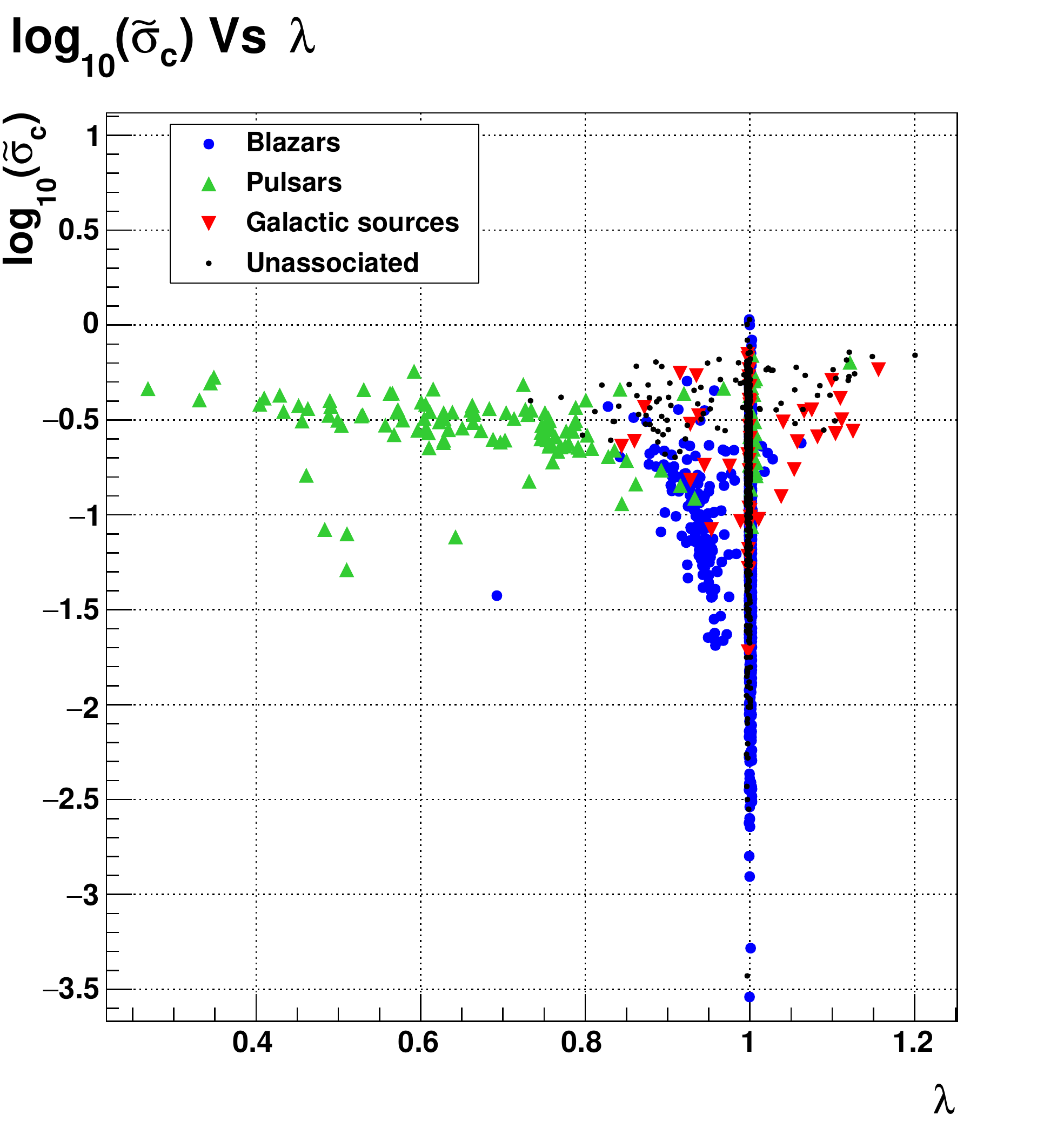}}
  \hspace{0.1\linewidth}
  \subfloat[]{
    \label{fig:HR23VsHR34}
    \includegraphics[width=0.75\columnwidth]{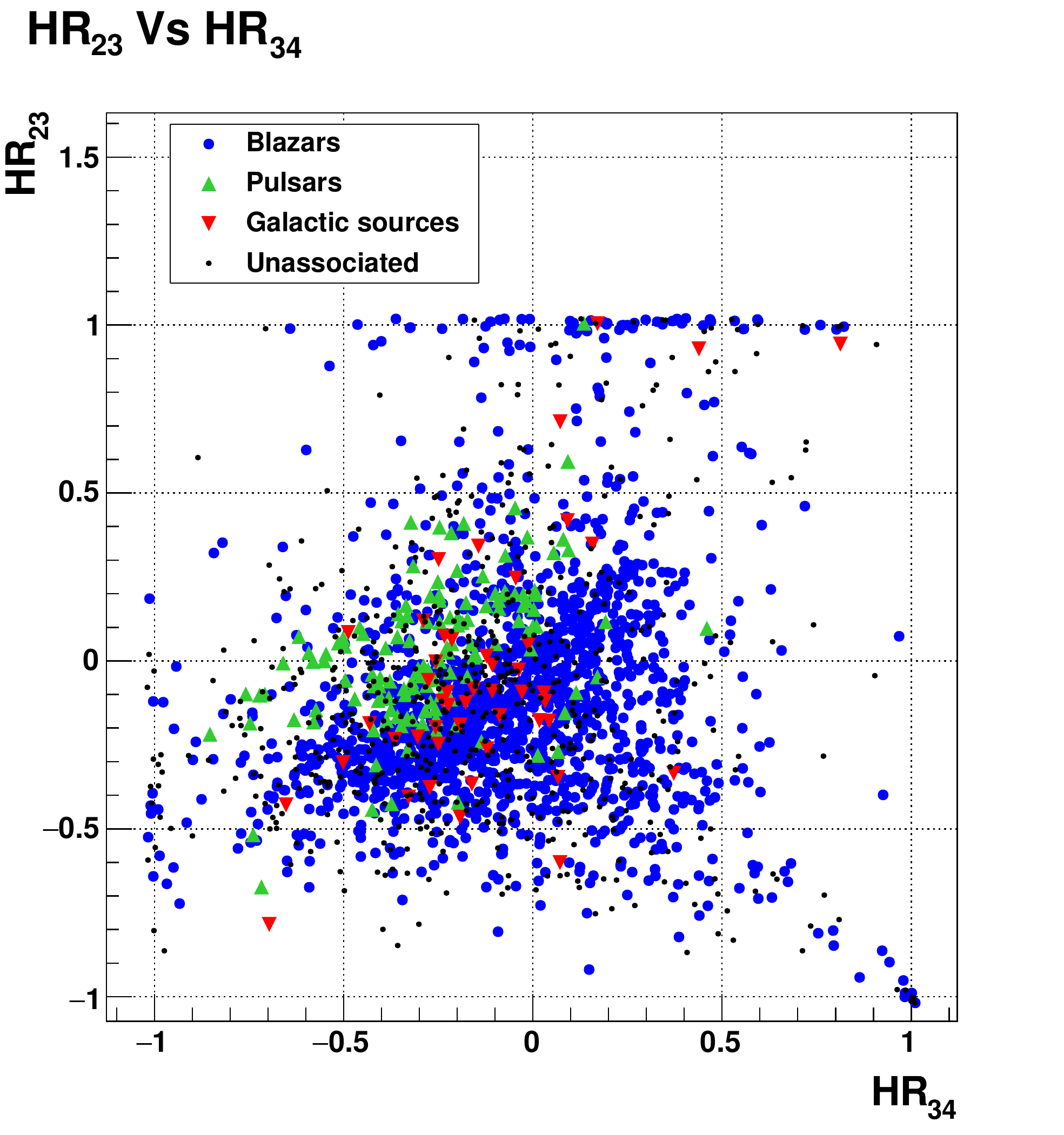}
  }\\[-2ex]

    \subfloat[]{
    \label{fig:LambdaVsCurvHR}
    \includegraphics[width=0.75\columnwidth]{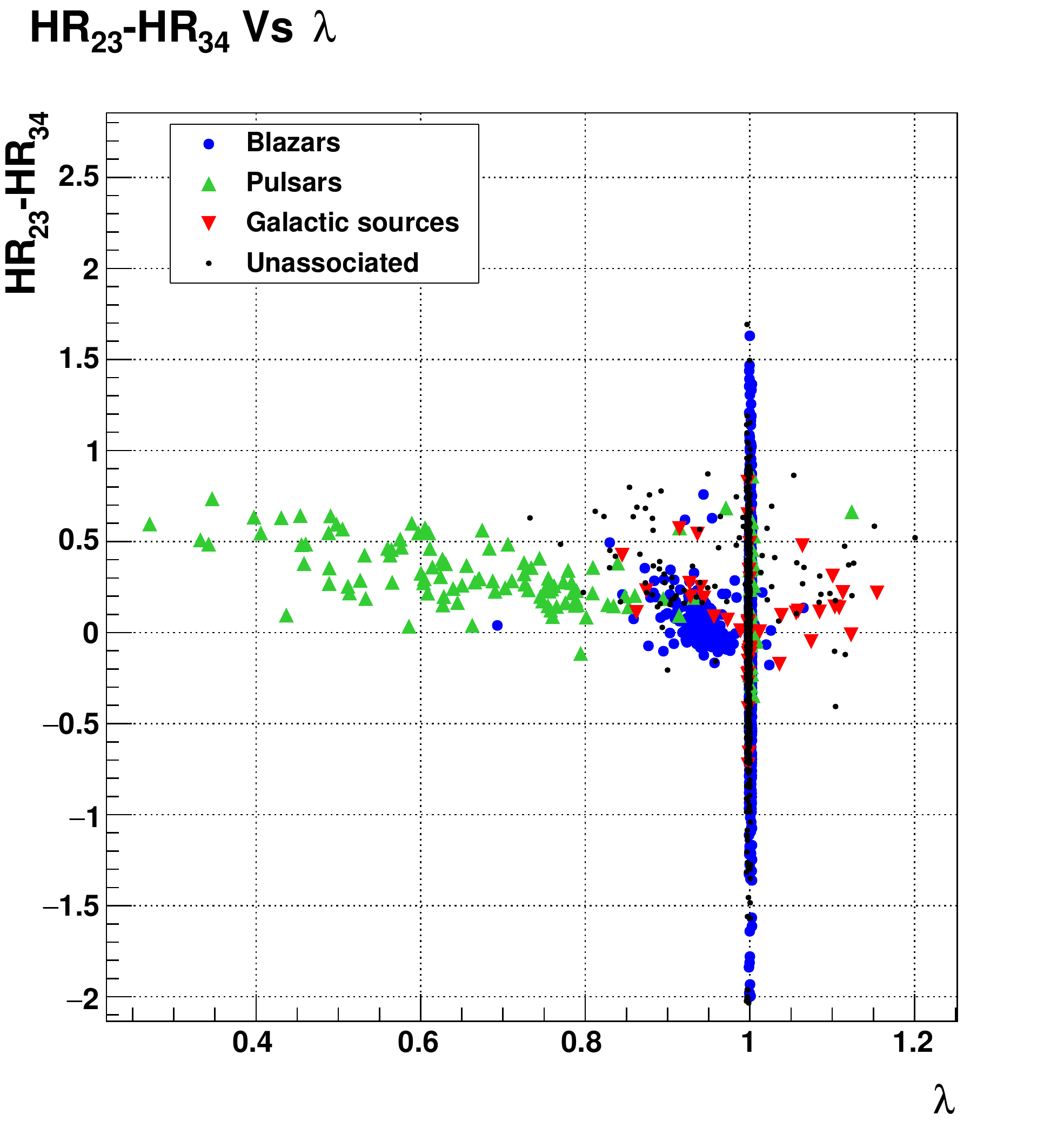}}
  \hspace{0.1\linewidth}
  \subfloat[] {
    \label{fig:NormCurvVsCurvHR}
    \includegraphics[width=0.75\columnwidth]{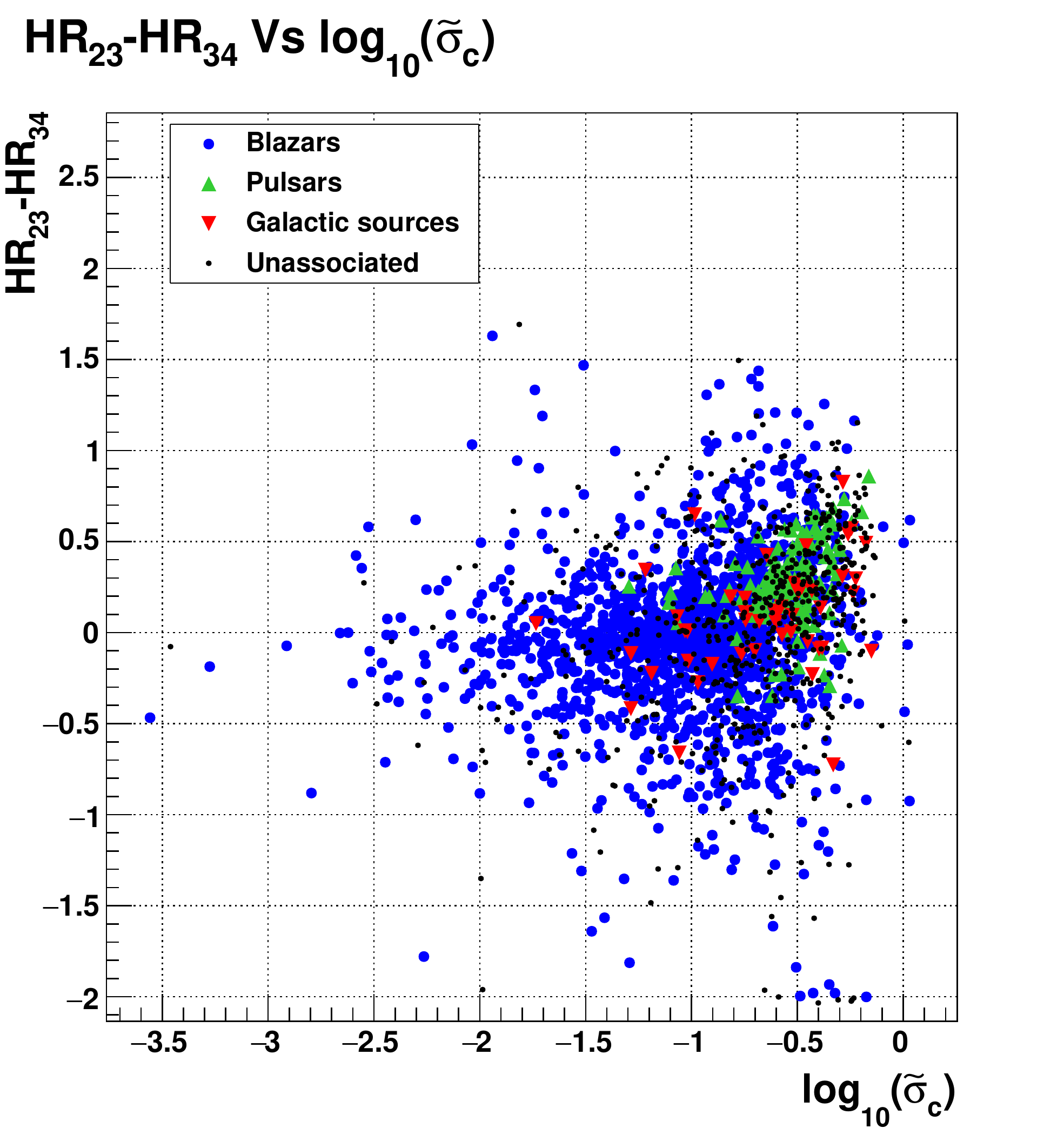}
  }
 
  \caption{Scatter plots for selected couples of discriminant parameters.
    Blazars are represented with blue circles, sources belonging to our Galaxy
    (except pulsars) with downward-pointing red triangles, pulsars with upward-pointing green triangles
    and unassociated sources with black dots.}
  \label{fig:FermiDistributions}

\end{figure*}

To distinguish between blazars and other source classes (case A), two types of parameters appear to be particularly powerful.
First, those quantifying the variability of the sources, which is a distinguishing feature of blazars over month-long time scales.
And second, spectral parameters, as blazar spectra are generally
well adjusted by a simple power law or a log parabola, whereas pulsars, for example, 
generally show a curved spectrum typically well adjusted by a broken power
law or a power law with an energy cut-off.
With this in mind, we reviewed the available parameters in the 3FGL catalogue 
and also examined those already used in previous
studies \citep{Ackermann:2012ab, Ferrara:2012aa, Mirabal2012, Doert:2013aa, Saz-Parkinson:2016aa}.
We finally selected six discriminant parameters, 
considering individually the increase of separation power and the stability
that they provide to the classifiers.
Five of these parameters have been used in previous studies: 
$\widetilde{\sigma}_c$, defined as $\sigma_c/\sigma$ where
$\sigma_c$ is the significance of the curvature and $\sigma$ is the
detection significance \citep{Doert:2013aa}; 
the normalised variability, called $\widetilde{\text{TS}}$, given
by the ratio between the index variability $\text{TS}$ and $\sigma$ \citep{Doert:2013aa}; 
and the hardness ratios\footnote{We use the definition of hardness ratio given in \citet{Ackermann:2012ab} which is 
 $\text{HR}_{ij} = \frac{ \Phi_j \langle E_j\rangle - \Phi_i \langle E_i\rangle}{
    \Phi_i \langle E_i\rangle +  \Phi_j \langle E_j\rangle}$, where $\Phi_i$ is the integral flux in the energy band $i$  
and $\langle E_i \rangle$ is the mean energy of the band.} 
$\text{HR}_{23}$ and $\text{HR}_{34}$ as well as their difference $\text{HR}_{23}-\text{HR}_{34}$
\citep{Ackermann:2012ab}.
We note that we chose to discard the hardness ratios
$\text{HR}_{12}$ and $\text{HR}_{45}$
in our selection for the lack of control of their discriminant
power\footnote{When a source is not
  detected in one of the five energy bands provided in the 3FGL catalogue,
  a $2\sigma$ upper limit
  is used by \citet{Acero:2015aa} instead of a flux measurement, leading to a shift of the hardness ratio determination.
  This is in particular the case of parameters $\text{HR}_{12}$ and especially $\text{HR}_{45}$,
  that have the bigger fractions of upper limits.}.
Additionally, we introduced a new parameter, called $\lambda$, defined as the ratio between
the spectral index of the preferred hypothesis and the
spectral index of the power law hypothesis, called $\gamma$.
Although for only ~17\% of sources in the 3FGL catalogue an alternative hypothesis is preferred over a power law,
this ratio increases to 76\% for pulsars while it is only ~9\% for blazars.
The distribution of $\lambda$ (when different from 1) shows an interesting separation power for blazars and pulsars,
see for example Fig.~\ref{fig:NormVarVsLambda}.
A selection of scatter plots is shown on Fig.~\ref{fig:FermiDistributions} for the selected set of discriminant parameters, 
considering the different source samples.

\begin{figure}
  \centering
  \subfloat[]{
    \label{fig:EnergyPivotVsGamma}
    \includegraphics[width=0.75\columnwidth]{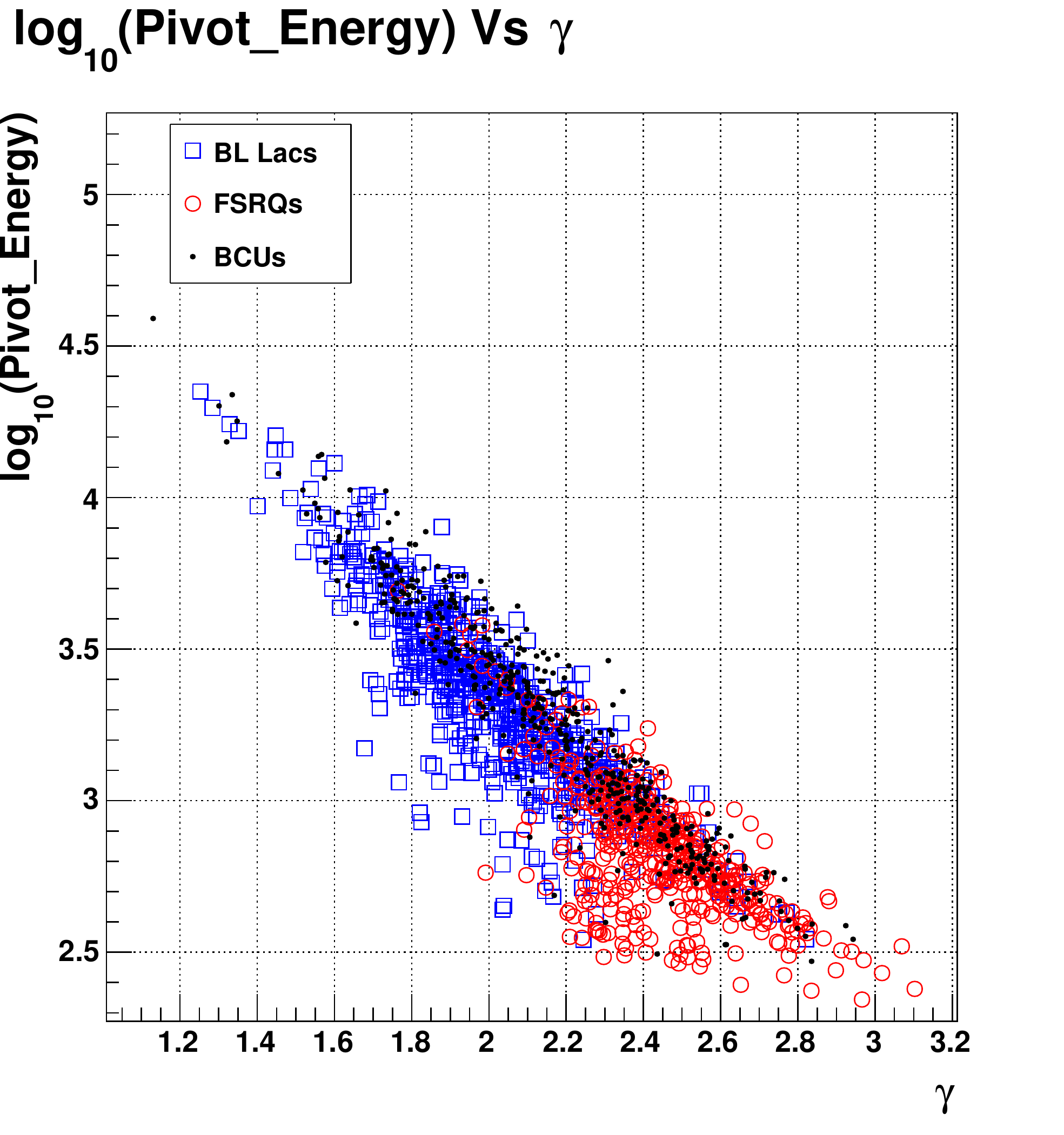}}\\[-0.ex]
  \subfloat[]{
    \label{fig:NormVariabilityVsGamma}        
    \includegraphics[width=0.75\columnwidth]{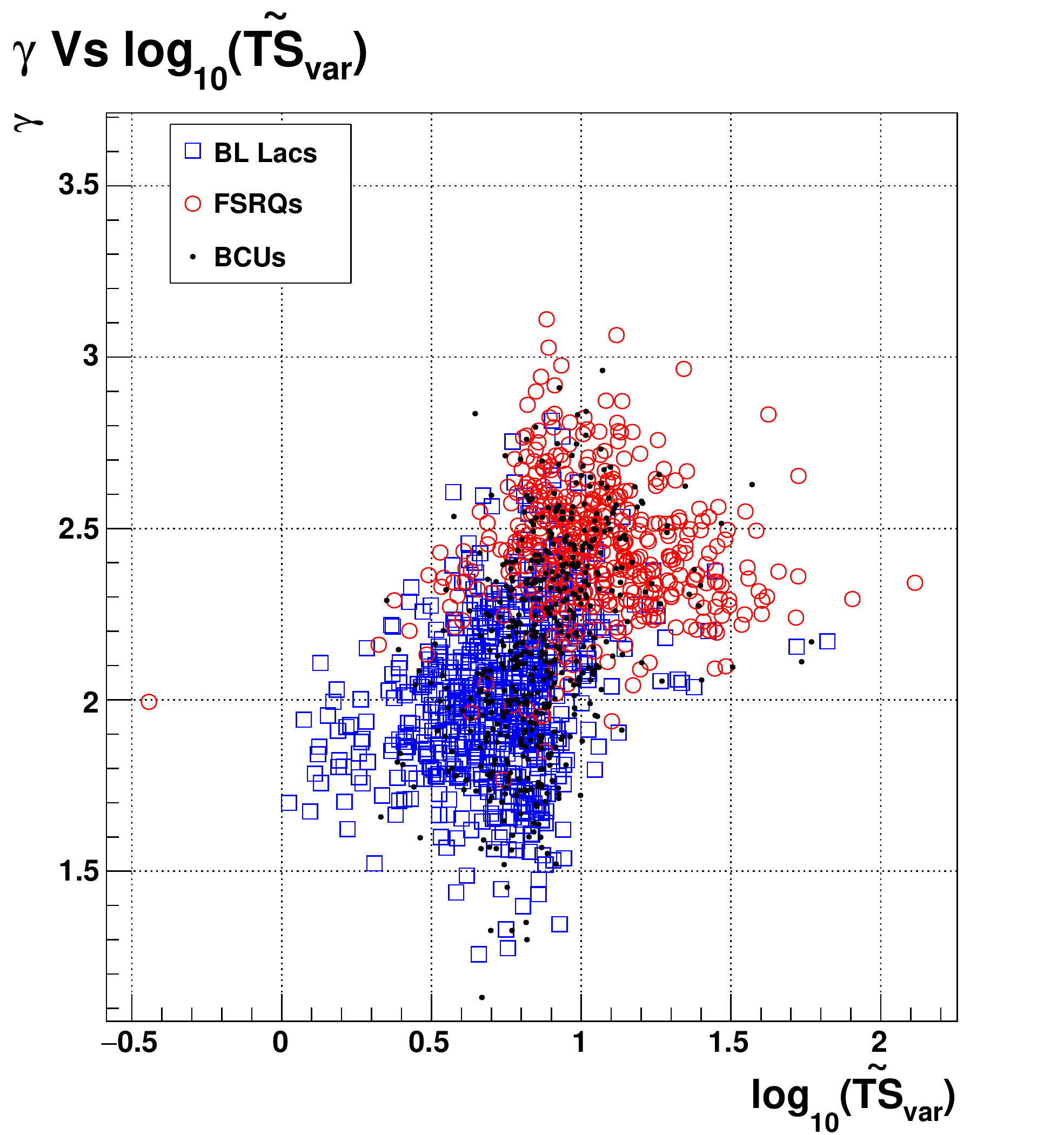}}\\[-0.ex]
  \subfloat[]{
    \label{fig:HR23VsHR34}
    \includegraphics[width=0.75\columnwidth]{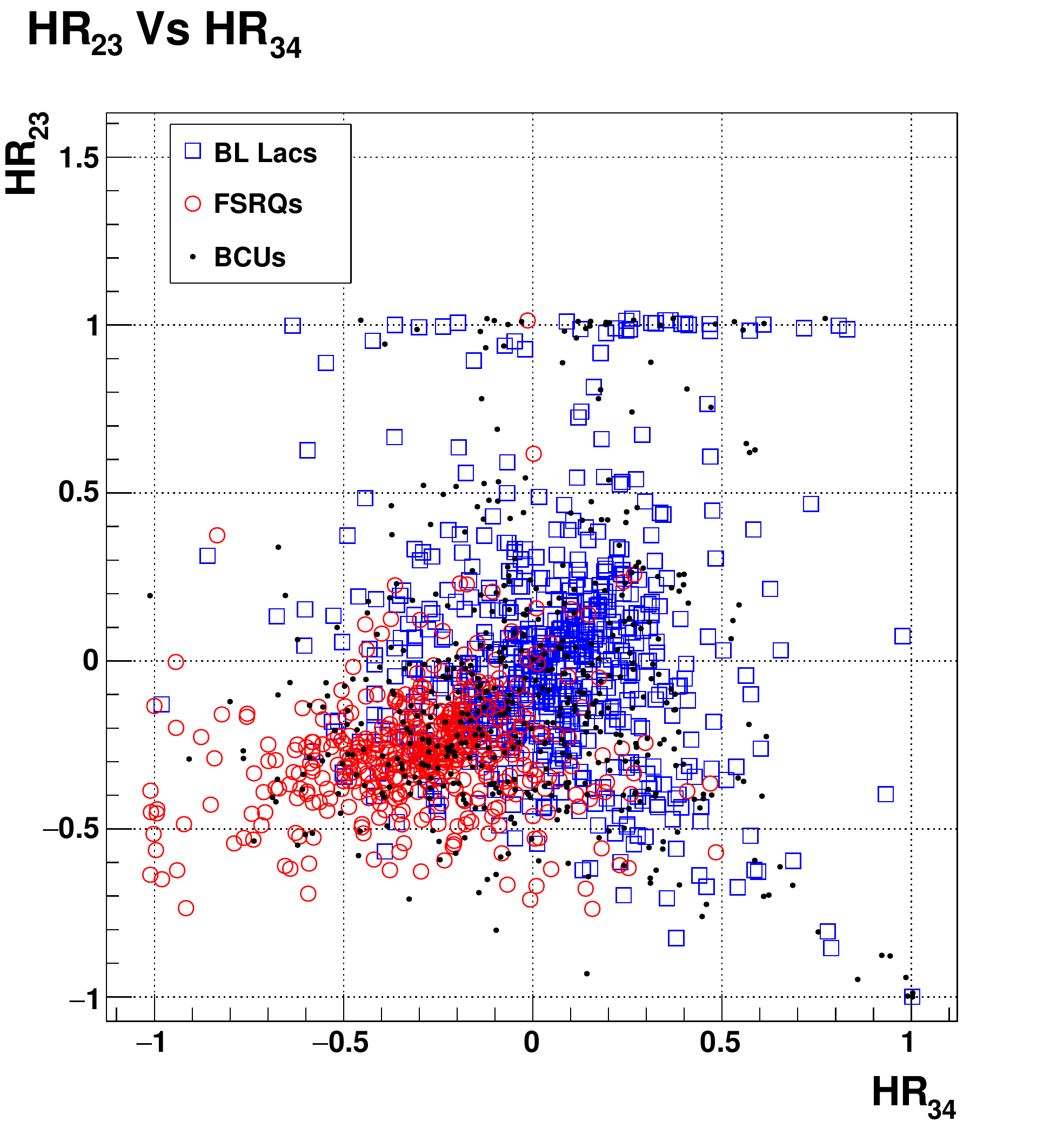}}
  \caption{Scatter plots for selected couples of discriminant parameters. \blls are represented with blue squares,
    \fsrqs with red circles and blazars of uncertain type with black dots.}
    \label{fig:FermiBllVsFsrqDistributions}
\end{figure}

The selection of a set of \bll/\fsrq discriminant parameters (case B) follows a similar approach.
The photon index $\gamma$, the pivot energy $E_{p}$ (which is
somewhat correlated to the position of the high energy peak) and 
the normalised variability $\widetilde{\text{TS}}$ were selected (Fig.~\ref{fig:FermiBllVsFsrqDistributions}
shows three scatter plots illustrating strong separation power). 
It is indeed shown in the Fermi/LAT 3LAC catalogue\footnote{The 3LAC catalogue is a by-product 
of the 3FGL catalogue devoted only to AGN sources.} \citep{Ackermann:2015aa},
that \fsrqs tend to have softer spectra than \blls, that their high energy peaks tend
to be located at lower energies, and that they tend to show stronger variability.
These parameters were also used in a similar study applied to the 2FGL catalogue
by \citet{Hassan:2013aa}. 
The set of six parameters selected above for the search of blazar
candidates among the unassociated 3FGL sources was also investigated.
In addition to $\widetilde{\text{TS}}$ which was already selected,
the hardness ratios $\text{HR}_{23}$ and $\text{HR}_{34}$
were also chosen. The other parameters were discarded,
  as they showed poor \bll/\fsrq separation power.


\section{Binary classifications based on machine-learning algorithms}
\label{Sec:FermiFermiMVA}

For this work, several machine-learning algorithms were tested 
in order to identify blazar candidates
among the 3FGL unassociated sources (case A) and also
to determine the \bll or \fsrq nature of blazars of unknown type (case B).
Using the Toolkit for Multivariate Data Analysis (\tmva) package \citep{Hoecker:2007aa}
it quickly appeared that, for a given set of discriminant parameters,
methods based on random forests, neural networks, support vector machines and
  boosted decision trees could reach comparable performance with very little tuning.
The choice was made to use two of these methods corresponding to different philosophies, the boosted
decision trees (BDT) and a multilayer perceptron (MLP) neural network.
In order to reduce the false association rate, the decisions of both 
classifiers were combined, then used to tag a source only if both
classifiers agree on its nature. 

The BDT machine-learning algorithm is based on decision trees,
a classifier structured on repeated yes/no decisions
designed to separate ``positive'' and ``negative'' classes of events.
Thereby, the phase space of the discriminant parameters is split
into two different regions.
The boosting algorithm, here AdaBoost \citep{Freund1996}, generates a forest of
weak decision trees and combines them to provide a final strong decision.
At each step, misclassified events are given an increasing
weight.
Then, the generation of the following tree is done with these weighted
events allowing the tree to become specialised on these difficult cases.
At the end of the boosting phase, new events can be processed by the forest of trees.
All decisions are then combined to give a
weighted response according to the specialisation of the trees.
Preliminary tests, performed in order to assess the stability of BDT classifiers,
have shown that similar performance was reached for a large range of BDT settings.
Considering this we decided to use the same settings for both case A and case B studies,
with values relatively close to those of the TMVA BDT default. Thus,
a large forest of short trees ($ntrees=400$, $depth=3$)
was generated with a learning rate of $0.2$.
The learning algorithm differs slightly from the original AdaBoost:
before the generation of a decision tree, during the boosting phase,
the events of the training samples are selected
$n$ times according to a given probability following a Poisson law of
parameter \SI{0.8}{} ($UseBaggedBoost=true$, $BaggedSampleFraction=0.8$).

Neural networks methods are based on artificial neurons.
It is possible to linearly separate two populations of events by building a
binary classifier with a single neuron.
The latter is composed of as many inputs as there are discriminant
parameters and one output describing the nature of the events.
To each input is associated a weight\footnote{There is generally an
  additional input called the bias to scale the output of the neuron.}.
Inside the neuron, using the weights, a linear combination of the discriminant
parameters is formed and then used as input for a transfer function which gives 
the output value of the neuron.
It is then possible to find the best values of the weights
allowing to get the minimum rate of misclassified events
by using a feedback process with the training sample.
Once this phase is finished, unknown events can be classified.
To tackle more complex problems, with non-linear separations between classes of events,
a possible solution is to use a
multilayer perceptron neural network \citep{Rumelhart1986}.
The latter is composed of at least one layer of neurons, called a hidden layer,
located between the input layer (made of as much single-input neurons as
there are discriminant parameters) and the output layer (made of a single neuron).
Additionally, each neuron is allowed to have direct connections with only the neurons
of the following layer.
The same procedure used for a single neuron is followed to adjust the weights.
As for the BDT classifiers, we found out that similar performance can be reached with a large range of MLP settings.
  We decided to use the same settings for both case A and B studies.
  We set the MLP architecture to a single
hidden-layer composed of $N_{\text{var}}+10$ neurons\footnote{$N_{\text{var}}$ is the number
of discriminant variables used to build up the classification.} and we used the back-propagation algorithm to find the minimum of the error function. 
Following the suggestion of \cite{Hoecker:2007aa}, the input variables
were normalised between $-1$ and $+1$ for the neural network.
Finally, as the positive and negative samples have different sizes, we normalised the events
in order to have samples with identical sizes\footnote{For BDT, this
  is naturally done with the AdaBoost algorithm.}
($NormMode=EqualNumEvents$).


\section{Training of classifiers and performance evaluation}
\label{Sec:TrainingAndPerf}

\subsection{Splitting labelled sources into training and test samples}
\label{Sec:SplitSamples}

A standard practice to build and evaluate a classifier is to create training and test samples by randomly
selecting, for example, \SI{70}{\percent} and \SI{30}{\percent} of each group of labelled events (here identified or associated sources).
The training sample is used to build the classifier, and the test sample to determine the performance metrics.
This random split is generally a good choice.
However in this work we had to handle with small data sets (sometimes composed of subsets of sources,
e.g. the galactic source sample for the case A study with 134 pulsars, 34 SNR or PWN, and 15 other galactic sources).
As shown by \citet{BrainWebb1999} this implies that the variance of classifiers corresponding to different randomly selected training subsamples
is likely to be important, leading to performance which could be significantly mis-estimated.

\begin{figure*}
  \centering
  \subfloat[]{
    \label{fig:superposed_roc_blazar_pulsar}
    \includegraphics[width=1.5\columnwidth]{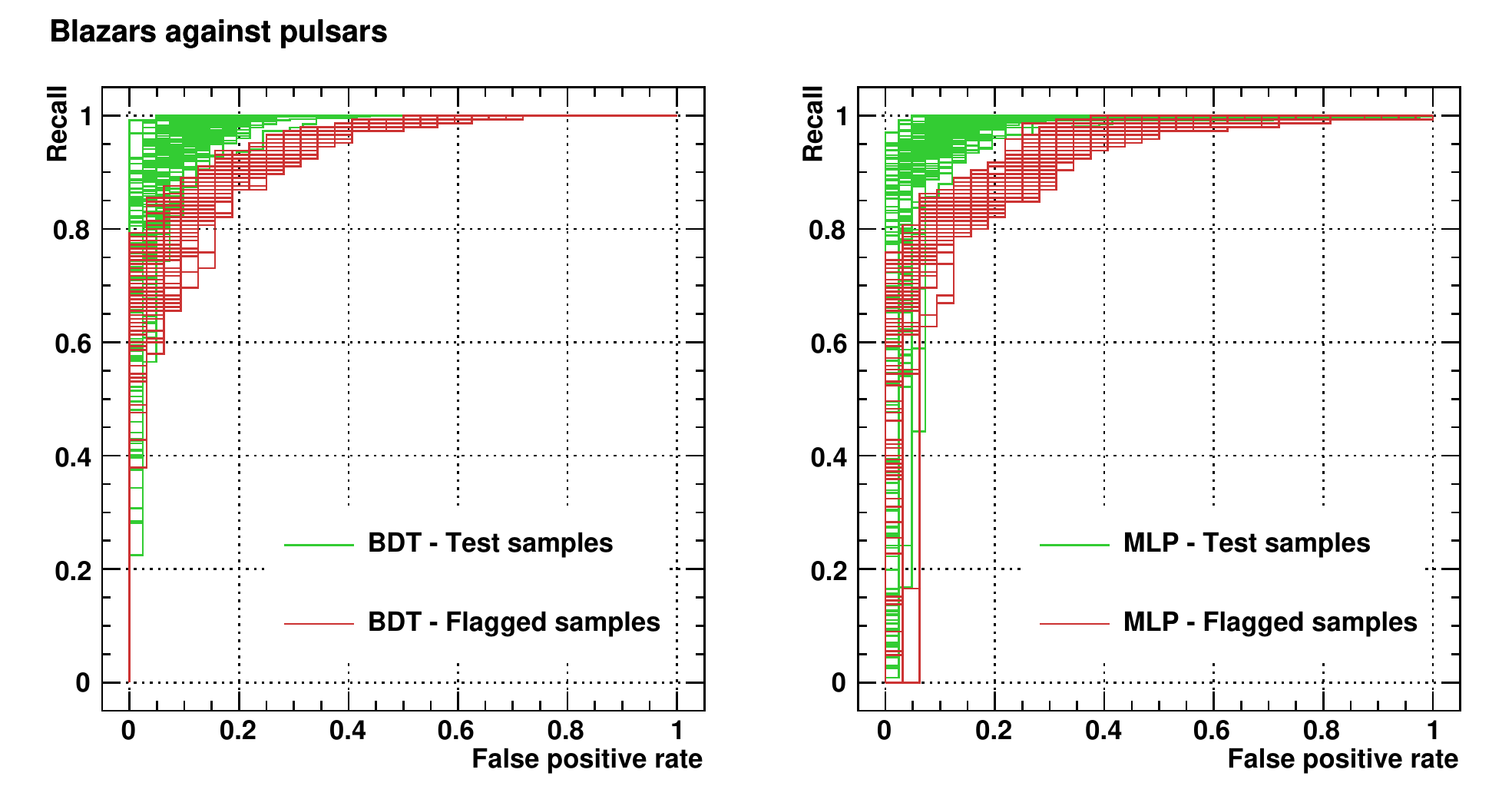}}\\[-0.2ex]
  \subfloat[]{
    \label{fig:superposed_roc_blazar_galactic}        
    \includegraphics[width=1.5\columnwidth]{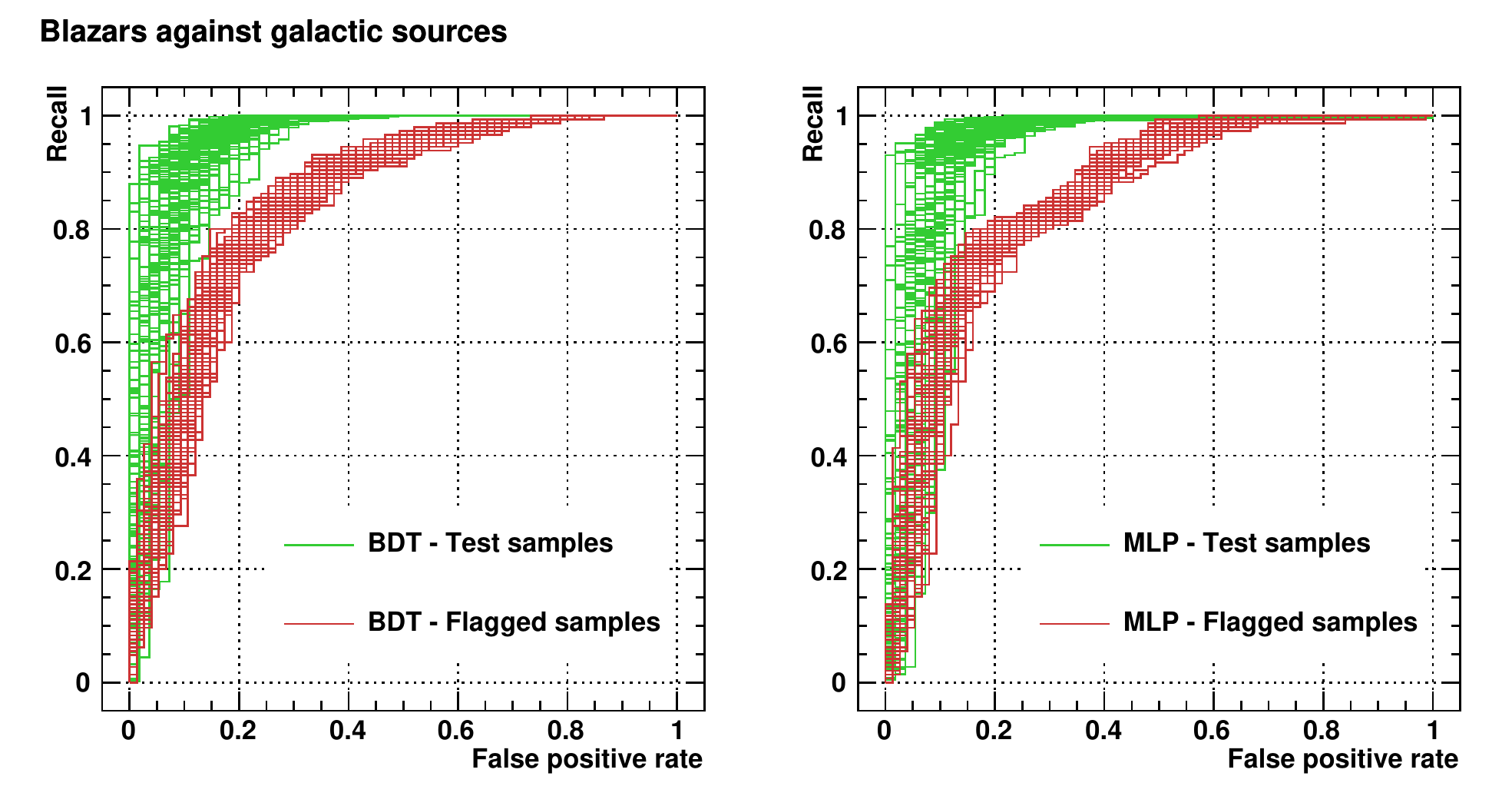}}\\[-0.2ex]
  \subfloat[]{
    \label{fig:superposed_roc_bll_fsrq}
    \includegraphics[width=1.5\columnwidth]{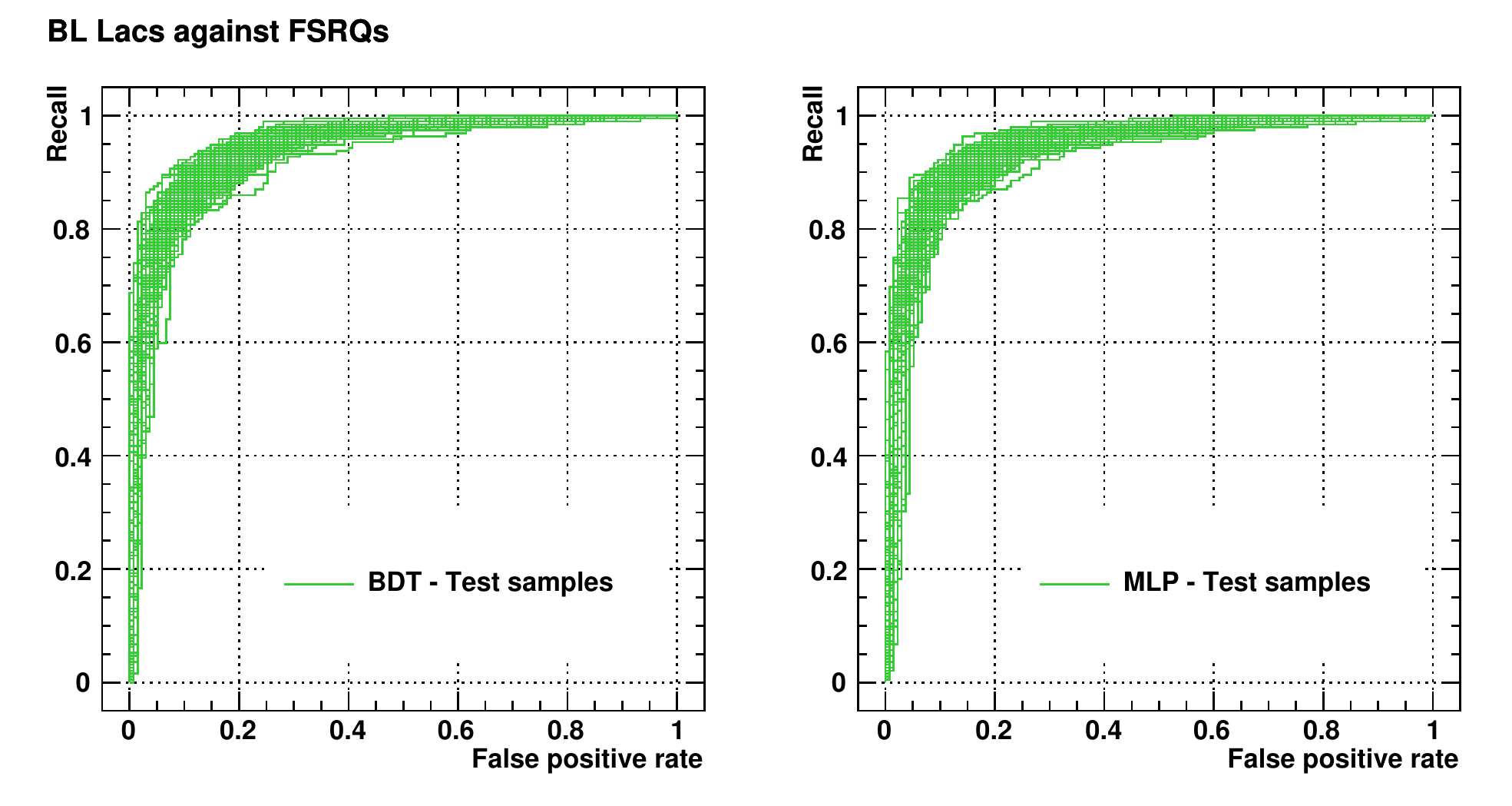}}
  \caption{ROC curves corresponding to 100 random splittings of the samples
    of labelled sources (with no flag) used for classifier building.
    Performance for sources with no flag were estimated using the test samples (green curves).
    Specific performance for flagged sources were estimated using the samples of labelled sources with a flag (red curves).
    The left and right columns show respectively the results for the BDT and MLP classifiers.
    Results for case A high galactic latitude, case A low galactic latitude and case B are shown in rows (a), (b) and (c), respectively.    
  }
  \label{fig:superposed_roc}
\end{figure*}

To minimise such a mis-estimation, we characterised the average performance of the BDT and MLP classifiers
(with respect to a large number of random splits of labelled sources into training and test samples),
and selected a single split which provides a pair of classifiers with performance as close as possible to this average behaviour.
To do that, we performed 100 iterations of the following sequence:
\begin{enumerate}
\item random split of the labelled samples in training  (\SI{70}{\percent}) and test (\SI{30}{\percent}) samples
\item training of the BDT and MLP methods using the same training sample
\item performance evaluation for BDT and MLP using the same test sample
\end{enumerate}
The receiver operating characteristic (ROC) curves\footnote{
  A ROC curve illustrates the performance of a classifier as its score threshold varies, representing the true positive rate against the false positive rate \citep{Fawcett:2006}.}
obtained for these 100 splittings are shown for case A (low and high galactic latitude) and B studies in Fig.~\ref{fig:superposed_roc}.
In each case, the training/test split which provides the performance closest to the average behaviour was selected ($\chi^2$ minimisation).

\subsection{Cutoff determination on the training sample}
\label{Sec:cutoffs}
The procedure explained above provided a pair of BDT and MLP classifiers for each study,
along with a training/test split of the sample of labelled sources.
To determine the optimal cutoff ($\zeta^{\star}$) in the distribution of the score ($\zeta$) generated by each classifier,
we used a ten-fold cross validation method on the training sample, following the sequence:
\begin{enumerate}
\item splitting of the training sample in ten equal-size subsamples.
\item training of the BDT and MLP classifiers on nine subsamples and application on the remaining $10^{\text{th}}$.
\item iteration over the ten subsamples until all the subsamples were tested.   
\item building of the BDT and MLP ROC curves on all the ten tested subsamples of the training sample.
\item determination of the $\zeta_{\text{BDT}}^{\star}$ and $\zeta_{\text{MLP}}^{\star}$ values considering a criterium defined below.
\end{enumerate}

A single criterium which ensures a low rate of false positives along with a relatively high
rate of true positives was used for all the studies (case A and B).
Our choice was to consider as cutoffs the values $\zeta_{\text{BDT}}^{\star}$ and $\zeta_{\text{MLP}}^{\star}$ which
provide for each classifier a false positive rate of \SI{10}{\percent}.
Consequently, for case B, two different cutoff values were obtained for the search of \blls or \fsrqs among
blazars (subsequently referred to as the \blls against \fsrqs or the \fsrqs against \blls studies, respectively).
All the cutoffs are summarised in Table~\ref{tab:FermiAnalysisPerf}.

\subsection{Performance metrics}
\label{Sec:Performance}

\begin{figure*}
  \centering
  \subfloat[]{
    \label{fig:finalperf_blazar_pulsar}
    \includegraphics[width=1.5\columnwidth]{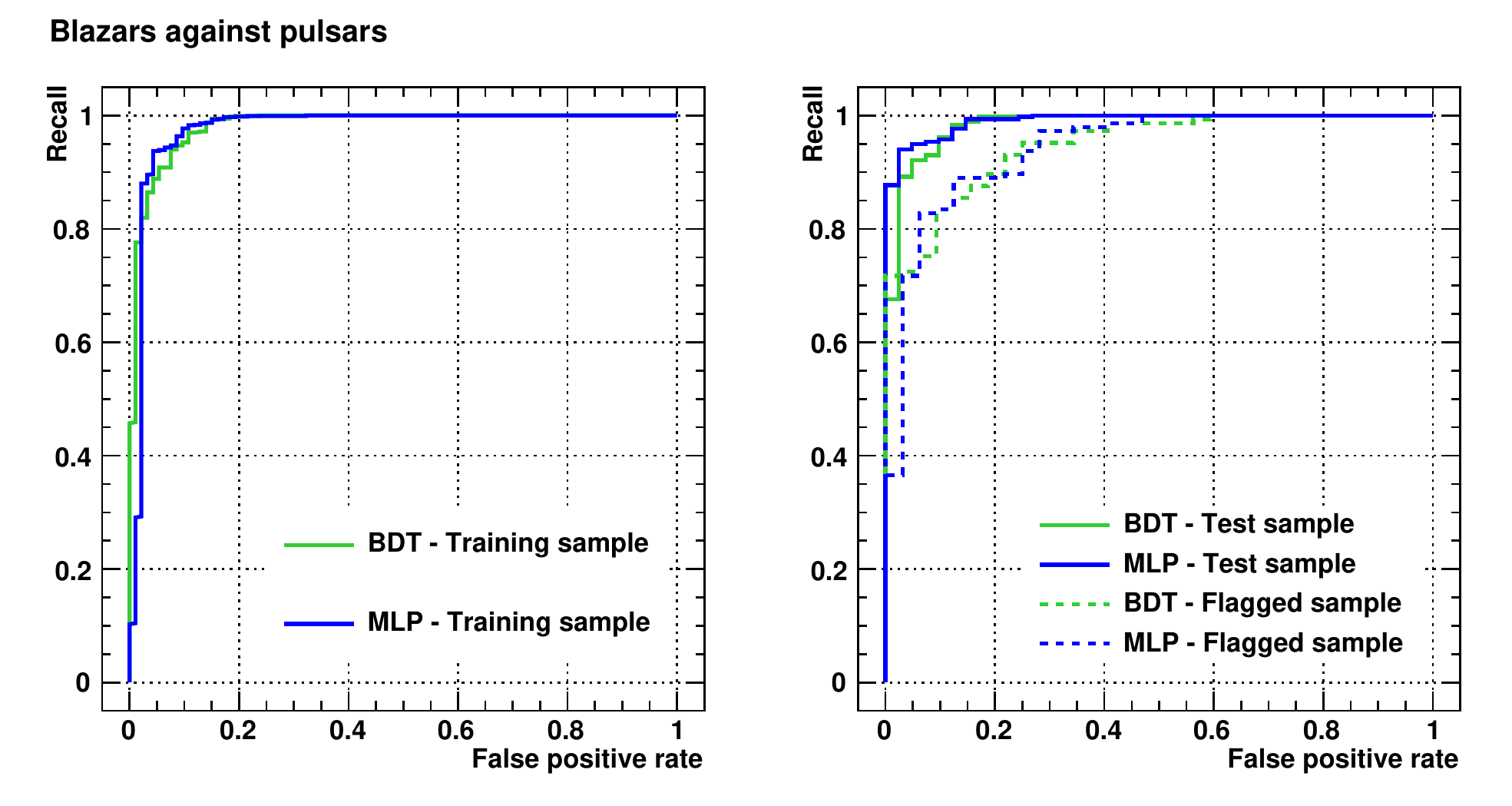}}\\[-0.2ex]
  \subfloat[]{
    \label{fig:finalperf_blazar_galactic}        
    \includegraphics[width=1.5\columnwidth]{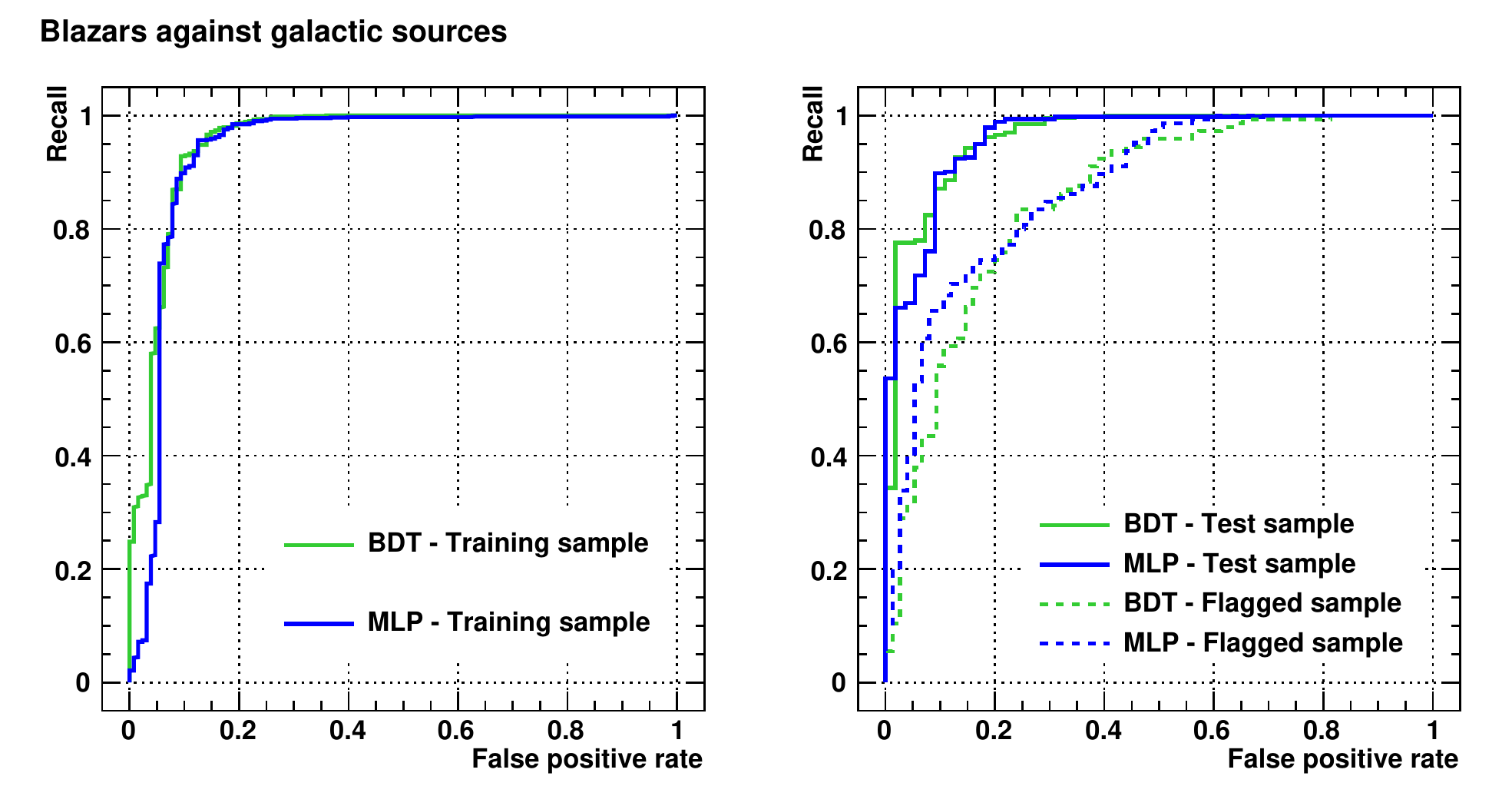}}\\[-0.2ex]
  \subfloat[]{
    \label{fig:finalperf_bll_fsrq}
    \includegraphics[width=1.5\columnwidth]{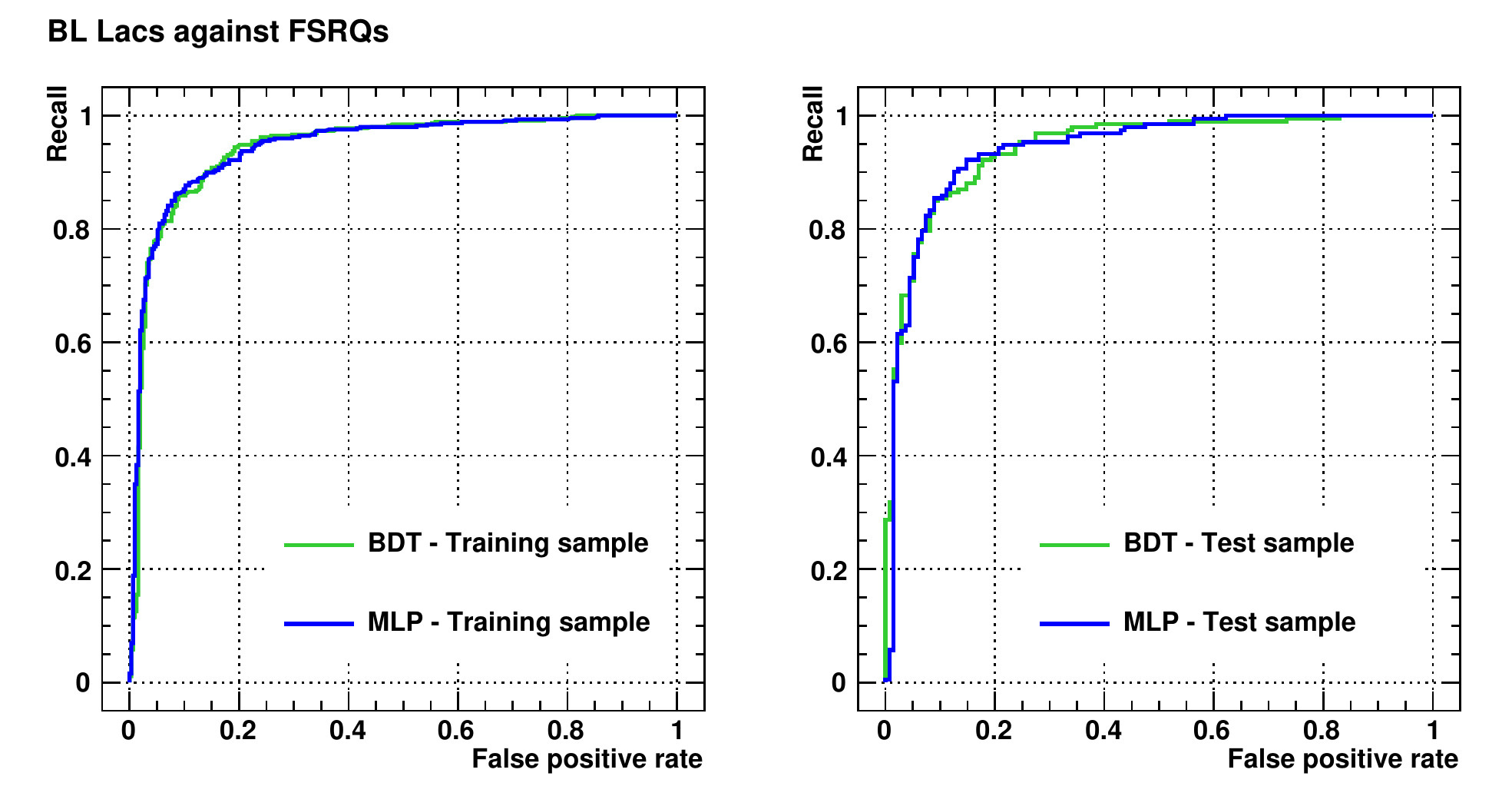}}
  \caption{ROC curves for classifiers
    used in case A high galactic latitude (a),
    case A low galactic latitude (b) and case B (c).
    In each case the left column shows the ROC curves (green for BDT, blue for MLP)
    obtained with the training sample using the ten-fold cross-validation method,
    as explained in Section~\ref{Sec:cutoffs}.
    The right column shows the ROC curves obtained when applying classifiers
    to the test sample (solid line) or to the sample of labelled sources with
    flag (dashed lines).}
  \label{fig:final_perf}
\end{figure*}

\begin{table*}
  \caption[]{Performance summary for the classifications of cases A  and B
    estimated on the test samples.
    For each of them we indicate the cutoff value $\zeta^{\star}$,
    the true positive (TP) rate and the false positive (FP) rate for the BDT
    and the MLP classifiers.
    The true positive rate and the false positive rate are also given when
    the decisions of the BDT and the MLP classifiers are combined together.}
  \centering
  \renewcommand{\tabcolsep}{2pt}
  \begin{tabular}{|c|l|c|c|c|c|c|c|c|c|}

    \hline

    \multirow{2}{*}{}
    & \multirow{2}{*}{Classifications}
    & \multicolumn{3}{c|}{BDT classifier}
    & \multicolumn{3}{c|}{MLP classifier}
    & \multicolumn{2}{c|}{Combination} \\ \cline{3-10}
    & & $\zeta^{\star}$
    & TP rate~(\SI{}{\percent})
    & FP rate~(\SI{}{\percent})
    & $\zeta^{\star}$
    & TP rate~(\SI{}{\percent})
    & FP rate~(\SI{}{\percent})
    & TP rate~(\SI{}{\percent})
    & FP rate~(\SI{}{\percent}) \\ \cline{1-10}

    \multirow{2}{*}{Case A}
    & Blazars against pulsars
    & \SI{0.408}{}
    & \SI{96.4}{}
    & \SI{12.2}{}
    & \SI{0.419}{}
    & \SI{97.5}{}
    & \SI{12.2}{}
    & \SI{95.6}{}
    & \SI{7.3}{} \\ \cline{2-10}
    & Blazars against galactic
    & \SI{0.320}{}
    & \SI{90.9}{}
    & \SI{12.7}{}
    & \SI{0.611}{}
    & \SI{88.7}{}
    & \SI{9.1}{}
    & \SI{87.1}{}
    & \SI{9.1}{} \\ \cline{1-10}

    \multirow{2}{*}{Case B}
    & \bll against \fsrq
    & \SI{0.076}{}
    & \SI{84.9}{}
    & \SI{9.6}{}
    & \SI{0.503}{}
    & \SI{85.9}{}
    & \SI{11.1}{}
    & \SI{83.9}{}
    & \SI{8.9}{} \\ \cline{2-10}
    & \fsrq against \bll
    & \SI{-0.007}{}
    & \SI{85.2}{}
    & \SI{13.0}{}
    & \SI{0.438}{}
    & \SI{87.4}{}
    & \SI{12.0}{}
    & \SI{84.4}{}
    & \SI{10.9}{} \\ \cline{1-10}

  \end{tabular}

  \label{tab:FermiAnalysisPerf}
\end{table*}

\begin{table*}

  \caption[]{Performance summary for the classifications of case A
    estimated on the flagged source samples.
    For each of them we indicate
    the true positive (TP) rate and the false positive (FP) rate for the BDT
    and the MLP classifiers.
    The true positive rate and the false positive rate are also given when
    the decisions of the BDT and the MLP classifiers are combined together.
    The cutoff values to derive those performance metrics
    are the same as the ones given in Table~\ref{tab:FermiAnalysisPerf}.}
  \centering
  \renewcommand{\tabcolsep}{2pt}
  \begin{tabular}{|c|l|c|c|c|c|c|c|}

    \hline

    \multirow{2}{*}{}
    & \multirow{2}{*}{Classifications}
    & \multicolumn{2}{c|}{BDT classifier}
    & \multicolumn{2}{c|}{MLP classifier}
    & \multicolumn{2}{c|}{Combination} \\ \cline{3-8}
    & & TP rate~(\SI{}{\percent})
    & FP rate~(\SI{}{\percent})
    & TP rate~(\SI{}{\percent})
    & FP rate~(\SI{}{\percent})
    & TP rate~(\SI{}{\percent})
    & FP rate~(\SI{}{\percent}) \\ \cline{1-8}

    \multirow{2}{*}{Case A}
    & Blazars against pulsars
    & \SI{91.0}{}
    & \SI{21.9}{}
    & \SI{91.0}{}
    & \SI{25.0}{}
    & \SI{88.9}{}
    & \SI{18.8}{} \\ \cline{2-8}
    & Blazars against galactic
    & \SI{87.6}{}
    & \SI{34.7}{}
    & \SI{84.1}{}
    & \SI{29.3}{}
    & \SI{81.4}{}
    & \SI{28.0}{} \\ \cline{1-8}

  \end{tabular} 

  \label{tab:FermiAnalysisPerfFlagged}
\end{table*}

Figure~\ref{fig:final_perf} shows the ROC curves for each study and each classifier,
first determined using the corresponding training sample (with the ten-fold cross-validation
method presented in Section~\ref{Sec:cutoffs}) and then using the test sample.
For the performance metrics evaluation, classifiers were applied to the test samples. We then used the cutoffs described above.
Combining the outputs of the BDT and MLP classifiers, we obtained a true positive rate of 95.6\% and a false positive rate of 7.3\%
for the blazars against pulsars study (case A, to be applied to high galactic latitude sources).
For the blazars against galactic sources study (case A, to be applied to low galactic latitude sources) we obtained slightly lower performance,
with a true positive rate of 87.1\% and a false positive rate of 9.1\%.
This loss of classifier performance is exclusively due to the inclusion of all galactic sources (in addition to pulsars) to the initial training sample.
For case B, similar performances were obtained for the \blls against \fsrqs and the \fsrqs against \blls studies, with true positive rates of 83.9\% and 84.4\% and false positive rates of
8.9\% and 10.9\%, respectively. 
All the true and false positive rates for the BDT and MLP classifiers (individual and combined) are summarised in Table~\ref{tab:FermiAnalysisPerf}.

As shown in Figs.~\ref{fig:finalperf_blazar_pulsar} and \ref{fig:finalperf_blazar_galactic}
(also visible in Figs.~\ref{fig:superposed_roc_blazar_pulsar} and \ref{fig:superposed_roc_blazar_galactic}),
the ROC curves obtained when applying the case A classifiers
to a sample of flagged sources are significantly different to the ones obtained to the test sample.
Consequently, considering the cutoff values obtained in Section~\ref{Sec:cutoffs}, we used the samples of flagged sources
to determine specific performance metrics for these categories of sources.
Combining the outputs of the BDT and MLP classifiers, we obtained true and false positive rates of 88.9\% and 18.8\% respectively
for the blazars against pulsars study, and true and false positive rates of 81.4\% and 28.0\% respectively
for the blazars against galactic sources study. In both cases, as compared to the performance obtained on non-flagged sources,
the true positive rates are slightly reduced (by \SI{\sim7}{\percent})
whereas the false positive rates significantly increase by a factor $\sim$2.6--3. The performance metrics for flagged sources are summarised in Table~\ref{tab:FermiAnalysisPerfFlagged}.

\section{Results}
\label{Sec:FermiFermiResults}

\begin{table}
  \centering
  \caption[]{Summary of results obtained when applying the classifiers to
    the high and low galactic latitude unassociated sources.
    $N_\text{s}$ and $N_\text{c}$ are the numbers of unassociated sources
    and blazar candidates, respectively.
    The number of false associations $N_\text{false}$  and the number of true blazars missed by the classifiers $N_\text{miss}$ are obtained
    considering that the $N_\text{s}$ unassociated sources are either blazars or galactic sources,
    and considering also the true and false positive rates given in Tables~\ref{tab:FermiAnalysisPerf} and \ref{tab:FermiAnalysisPerfFlagged}.
   }
  \renewcommand{\tabcolsep}{2pt}
  \begin{tabular}{|c||c|c|c|c|c|}

    \hline
    \multicolumn{1}{|c|}{Target}
    & \multicolumn{1}{|c}{Flag}
    & \multicolumn{1}{|c}{$N_\text{s}$}
    & \multicolumn{1}{|c}{$N_\text{c}$}
    & \multicolumn{1}{|c|}{$N_\text{false}$} 
    & \multicolumn{1}{|c|}{$N_\text{miss}$} \\

    \hline

    \multicolumn{1}{|c|}{High-latitude}
    & \multicolumn{1}{|l}{no}
    & \multicolumn{1}{|c}{422}
    & \multicolumn{1}{|c}{345}
    & \multicolumn{1}{|c|}{\SI{4.8}{}} 
    & \multicolumn{1}{|c|}{\SI{15.8}{}} \\

    \cline{2-6}

    \multicolumn{1}{|c|}{unassociated sources}
    & \multicolumn{1}{|l}{yes}
    & \multicolumn{1}{|c}{109}
    & \multicolumn{1}{|c}{80}
    & \multicolumn{1}{|c|}{\SI{4.5}} 
    & \multicolumn{1}{|c|}{\SI{9.4}{}} \\

    \hline

    \hline

    \multicolumn{1}{|c|}{Low-latitude}
    & \multicolumn{1}{|l}{no}
    & \multicolumn{1}{|c}{169}
    & \multicolumn{1}{|c}{72}
    & \multicolumn{1}{|c|}{\SI{8.8}} 
    & \multicolumn{1}{|c|}{\SI{9.4}{}} \\

    \cline{2-6}

    \multicolumn{1}{|c|}{unassociated sources}
    & \multicolumn{1}{|l}{yes}
    & \multicolumn{1}{|c}{247}
    & \multicolumn{1}{|c}{98}
    & \multicolumn{1}{|c|}{\SI{54.0}{}} 
    & \multicolumn{1}{|c|}{\SI{10.1}{}} \\

    \hline

  \end{tabular}

  \label{tab:FermiAnalysisResults}
\end{table}

The results presented below were obtained by combining the BDT and MLP decisions for each study.

For the blazars against pulsars study (case A, to be applied to high galactic latitude sources),
the classifiers were applied to 531 unassociated sources (422 not flagged, 109 flagged) with $|b| > \SI{10}{\degree}$.
This results in 425 blazar candidates (345 not flagged, 80 flagged) with the number of false associations estimated to $\sim$9.3
(4.8 not flagged and 4.5 flagged).
For the blazars against galactic sources study (case A, to be applied to low galactic latitude sources),
the classifiers were applied to 416 unassociated sources (169 not flagged, 247 flagged) with $|b| \leq \SI{10}{\degree}$.
This results in 72 blazar candidates among the 169 unassociated sources with no flag, with the number of false associations estimated to be approximately nine.
In addition we obtained 98 blazar candidates among the 247 unassociated sources with a flag, but this sample is dominated by false associations,
which are estimated to be $\sim$54.  
Results are summarised in Table~\ref{tab:FermiAnalysisResults} and a short sample of sources is shown in Table~\ref{tab:FermiAnalysisTable}.

\tiny
\renewcommand{\tabcolsep}{2pt}
\begin{table*}
  \caption{Example illustrating the structure of tables provided in 
    \protect\href{https://unidgamma.in2p3.fr}{https://unidgamma.in2p3.fr} as an output of the case A study.
    Here we show the sample of the twenty brightest blazar candidates with
    no flag in the high galactic latitude region ($|b| > \SI{10}{\degree}$). 
    The columns correspond respectively to the 3FGL source name,
    the galactic coordinates of the source ($l$,$b$), 
    the values of the six discriminant parameters, and the values
    of the output parameters $\zeta$ built with the BDT and MLP classifiers.
    We added in the last column the assignation from the case B study for the \bll or \fsrq nature of the source 
    (bll for \bll, fsrq for \fsrq and unc for uncertain).
    The superscripts correspond to a blazar candidate previously proposed by other authors:
      $\dagger$ for \citet{Saz-Parkinson:2016aa},  $\star$ for \citet{Doert:2013aa}, $\sharp$ for \citet{Mirabal2012},
      $\circ$ for \citet{Massaro:2013aa} and $\diamond$ for \citet{Ackermann:2012ab}.
  }

  \centering

  \begin{tabular}{|l||c|c|c|c|c|c|c|c|c|c|c|}

    \hline

    \multicolumn{1}{|l||}{3FGL name}
    & \multicolumn{1}{c|}{$l$ (\SI{}{\degree})}
    & \multicolumn{1}{c|}{$b$ (\SI{}{\degree})}
    & \multicolumn{1}{c|}{$\log_{10} \widetilde{\sigma}_c$}
    & \multicolumn{1}{c|}{$\log_{10} \widetilde{\text{TS}}$}
    & \multicolumn{1}{c|}{$\text{HR}_{23}$}
    & \multicolumn{1}{c|}{$\text{HR}_{34}$}
    & \multicolumn{1}{c|}{$\text{HR}_{23}-\text{HR}_{34}$}
    & \multicolumn{1}{c|}{$\lambda$}
    & \multicolumn{1}{c|}{$\zeta_{\text{BDT}}$}
    & \multicolumn{1}{c|}{$\zeta_{\text{MLP}}$} 
    & \multicolumn{1}{c|}{type} \\

    \hline
    \hline

    \multicolumn{1}{|l||}{J0032.3-5522$^{\dagger}$$^{\star}$$^{\sharp}$$^{\diamond}$} & \multicolumn{1}{c|}{308.619} & \multicolumn{1}{c|}{-61.549} & \multicolumn{1}{c|}{-1.12} & \multicolumn{1}{c|}{1.45} & \multicolumn{1}{c|}{-0.24} & \multicolumn{1}{c|}{-0.21} & \multicolumn{1}{c|}{-0.03} & \multicolumn{1}{c|}{1.00} & \multicolumn{1}{c|}{0.8563} & \multicolumn{1}{c|}{0.9955} & \multicolumn{1}{c|}{fsrq} \\ 
\hline 
\multicolumn{1}{|l||}{J0537.0+0957$^{\dagger}$} & \multicolumn{1}{c|}{195.285} & \multicolumn{1}{c|}{-11.578} & \multicolumn{1}{c|}{-0.42} & \multicolumn{1}{c|}{1.20} & \multicolumn{1}{c|}{-0.50} & \multicolumn{1}{c|}{-0.69} & \multicolumn{1}{c|}{0.18} & \multicolumn{1}{c|}{0.89} & \multicolumn{1}{c|}{0.5999} & \multicolumn{1}{c|}{0.9299} & \multicolumn{1}{c|}{fsrq} \\ 
\hline 
\multicolumn{1}{|l||}{J0940.6-7609} & \multicolumn{1}{c|}{292.247} & \multicolumn{1}{c|}{-17.425} & \multicolumn{1}{c|}{-0.95} & \multicolumn{1}{c|}{0.68} & \multicolumn{1}{c|}{0.08} & \multicolumn{1}{c|}{-0.48} & \multicolumn{1}{c|}{0.56} & \multicolumn{1}{c|}{1.00} & \multicolumn{1}{c|}{0.6433} & \multicolumn{1}{c|}{0.6872} & \multicolumn{1}{c|}{unc} \\ 
\hline 
\multicolumn{1}{|l||}{J1050.4+0435$^{\dagger}$} & \multicolumn{1}{c|}{245.550} & \multicolumn{1}{c|}{53.413} & \multicolumn{1}{c|}{-0.80} & \multicolumn{1}{c|}{1.34} & \multicolumn{1}{c|}{-0.22} & \multicolumn{1}{c|}{-0.45} & \multicolumn{1}{c|}{0.23} & \multicolumn{1}{c|}{1.00} & \multicolumn{1}{c|}{0.8447} & \multicolumn{1}{c|}{1.2406} & \multicolumn{1}{c|}{fsrq} \\ 
\hline 
\multicolumn{1}{|l||}{J1221.5-0632$^{\dagger}$$^{\star}$$^{\sharp}$$^{\diamond}$} & \multicolumn{1}{c|}{289.713} & \multicolumn{1}{c|}{55.552} & \multicolumn{1}{c|}{-0.75} & \multicolumn{1}{c|}{0.66} & \multicolumn{1}{c|}{0.04} & \multicolumn{1}{c|}{-0.07} & \multicolumn{1}{c|}{0.10} & \multicolumn{1}{c|}{1.00} & \multicolumn{1}{c|}{0.5249} & \multicolumn{1}{c|}{0.6722} & \multicolumn{1}{c|}{bll} \\ 
\hline 
\multicolumn{1}{|l||}{J1315.7-0732$^{\dagger}$$^{\star}$$^{\sharp}$$^{\circ}$$^{\diamond}$} & \multicolumn{1}{c|}{313.448} & \multicolumn{1}{c|}{54.825} & \multicolumn{1}{c|}{-1.82} & \multicolumn{1}{c|}{0.75} & \multicolumn{1}{c|}{0.05} & \multicolumn{1}{c|}{-0.24} & \multicolumn{1}{c|}{0.29} & \multicolumn{1}{c|}{1.00} & \multicolumn{1}{c|}{0.6916} & \multicolumn{1}{c|}{1.0256} & \multicolumn{1}{c|}{bll} \\ 
\hline 
\multicolumn{1}{|l||}{J1335.2-4056$^{\dagger}$$^{\star}$$^{\sharp}$} & \multicolumn{1}{c|}{311.784} & \multicolumn{1}{c|}{21.182} & \multicolumn{1}{c|}{-0.77} & \multicolumn{1}{c|}{0.85} & \multicolumn{1}{c|}{-0.49} & \multicolumn{1}{c|}{-0.27} & \multicolumn{1}{c|}{-0.22} & \multicolumn{1}{c|}{1.00} & \multicolumn{1}{c|}{0.5341} & \multicolumn{1}{c|}{0.8333} & \multicolumn{1}{c|}{fsrq} \\ 
\hline 
\multicolumn{1}{|l||}{J1417.5-4402$^{\dagger}$$^{\star}$$^{\sharp}$$^{\diamond}$} & \multicolumn{1}{c|}{318.864} & \multicolumn{1}{c|}{16.150} & \multicolumn{1}{c|}{-0.93} & \multicolumn{1}{c|}{0.64} & \multicolumn{1}{c|}{-0.10} & \multicolumn{1}{c|}{-0.09} & \multicolumn{1}{c|}{-0.01} & \multicolumn{1}{c|}{1.00} & \multicolumn{1}{c|}{0.7065} & \multicolumn{1}{c|}{0.8152} & \multicolumn{1}{c|}{unc} \\ 
\hline 
\multicolumn{1}{|l||}{J1417.7-5026$^{\dagger}$$^{\star}$$^{\sharp}$} & \multicolumn{1}{c|}{316.692} & \multicolumn{1}{c|}{10.113} & \multicolumn{1}{c|}{-1.19} & \multicolumn{1}{c|}{0.82} & \multicolumn{1}{c|}{-0.03} & \multicolumn{1}{c|}{-0.36} & \multicolumn{1}{c|}{0.33} & \multicolumn{1}{c|}{1.00} & \multicolumn{1}{c|}{0.6654} & \multicolumn{1}{c|}{0.9688} & \multicolumn{1}{c|}{fsrq} \\ 
\hline 
\multicolumn{1}{|l||}{J1548.4+1455$^{\dagger}$$^{\star}$$^{\sharp}$} & \multicolumn{1}{c|}{25.633} & \multicolumn{1}{c|}{47.175} & \multicolumn{1}{c|}{-1.05} & \multicolumn{1}{c|}{0.59} & \multicolumn{1}{c|}{-0.01} & \multicolumn{1}{c|}{-0.13} & \multicolumn{1}{c|}{0.11} & \multicolumn{1}{c|}{1.00} & \multicolumn{1}{c|}{0.6562} & \multicolumn{1}{c|}{0.8915} & \multicolumn{1}{c|}{bll} \\ 
\hline 
\multicolumn{1}{|l||}{J1625.6-2058$^{\dagger}$} & \multicolumn{1}{c|}{355.668} & \multicolumn{1}{c|}{19.288} & \multicolumn{1}{c|}{-0.68} & \multicolumn{1}{c|}{0.78} & \multicolumn{1}{c|}{-0.65} & \multicolumn{1}{c|}{-0.08} & \multicolumn{1}{c|}{-0.57} & \multicolumn{1}{c|}{1.00} & \multicolumn{1}{c|}{0.6440} & \multicolumn{1}{c|}{0.7777} & \multicolumn{1}{c|}{fsrq} \\ 
\hline 
\multicolumn{1}{|l||}{J1704.1+1234$^{\dagger}$$^{\star}$$^{\sharp}$} & \multicolumn{1}{c|}{32.465} & \multicolumn{1}{c|}{29.424} & \multicolumn{1}{c|}{-1.30} & \multicolumn{1}{c|}{0.70} & \multicolumn{1}{c|}{-0.42} & \multicolumn{1}{c|}{-0.13} & \multicolumn{1}{c|}{-0.30} & \multicolumn{1}{c|}{1.00} & \multicolumn{1}{c|}{0.7970} & \multicolumn{1}{c|}{0.9701} & \multicolumn{1}{c|}{unc} \\ 
\hline 
\multicolumn{1}{|l||}{J1704.4-0528$^{\dagger}$$^{\star}$$^{\sharp}$} & \multicolumn{1}{c|}{14.913} & \multicolumn{1}{c|}{20.797} & \multicolumn{1}{c|}{-1.34} & \multicolumn{1}{c|}{0.72} & \multicolumn{1}{c|}{-0.53} & \multicolumn{1}{c|}{0.34} & \multicolumn{1}{c|}{-0.86} & \multicolumn{1}{c|}{1.00} & \multicolumn{1}{c|}{0.8318} & \multicolumn{1}{c|}{0.9764} & \multicolumn{1}{c|}{bll} \\ 
\hline 
\multicolumn{1}{|l||}{J1747.3+0324$^{\star}$$^{\sharp}$} & \multicolumn{1}{c|}{28.606} & \multicolumn{1}{c|}{15.797} & \multicolumn{1}{c|}{-0.74} & \multicolumn{1}{c|}{0.81} & \multicolumn{1}{c|}{-0.05} & \multicolumn{1}{c|}{-0.04} & \multicolumn{1}{c|}{-0.01} & \multicolumn{1}{c|}{1.00} & \multicolumn{1}{c|}{0.6598} & \multicolumn{1}{c|}{0.8106} & \multicolumn{1}{c|}{bll} \\ 
\hline 
\multicolumn{1}{|l||}{J1801.5-7825$^{\dagger}$} & \multicolumn{1}{c|}{315.325} & \multicolumn{1}{c|}{-24.088} & \multicolumn{1}{c|}{-1.31} & \multicolumn{1}{c|}{1.01} & \multicolumn{1}{c|}{-0.32} & \multicolumn{1}{c|}{-0.28} & \multicolumn{1}{c|}{-0.05} & \multicolumn{1}{c|}{1.00} & \multicolumn{1}{c|}{0.8304} & \multicolumn{1}{c|}{0.9857} & \multicolumn{1}{c|}{fsrq} \\ 
\hline 
\multicolumn{1}{|l||}{J1816.0-6407} & \multicolumn{1}{c|}{330.337} & \multicolumn{1}{c|}{-20.519} & \multicolumn{1}{c|}{-0.70} & \multicolumn{1}{c|}{0.71} & \multicolumn{1}{c|}{-0.27} & \multicolumn{1}{c|}{-0.32} & \multicolumn{1}{c|}{0.04} & \multicolumn{1}{c|}{1.00} & \multicolumn{1}{c|}{0.4172} & \multicolumn{1}{c|}{0.4639} & \multicolumn{1}{c|}{fsrq} \\ 
\hline 
\multicolumn{1}{|l||}{J1845.5-2524} & \multicolumn{1}{c|}{9.441} & \multicolumn{1}{c|}{-10.079} & \multicolumn{1}{c|}{-0.69} & \multicolumn{1}{c|}{0.64} & \multicolumn{1}{c|}{-0.35} & \multicolumn{1}{c|}{0.20} & \multicolumn{1}{c|}{-0.55} & \multicolumn{1}{c|}{1.00} & \multicolumn{1}{c|}{0.4874} & \multicolumn{1}{c|}{0.7558} & \multicolumn{1}{c|}{bll} \\ 
\hline 
\multicolumn{1}{|l||}{J1917.1-3024$^{\dagger}$$^{\star}$$^{\sharp}$$^{\diamond}$} & \multicolumn{1}{c|}{7.551} & \multicolumn{1}{c|}{-18.469} & \multicolumn{1}{c|}{-1.38} & \multicolumn{1}{c|}{0.68} & \multicolumn{1}{c|}{-0.09} & \multicolumn{1}{c|}{0.04} & \multicolumn{1}{c|}{-0.13} & \multicolumn{1}{c|}{1.00} & \multicolumn{1}{c|}{0.8109} & \multicolumn{1}{c|}{1.0398} & \multicolumn{1}{c|}{bll} \\ 
\hline 
\multicolumn{1}{|l||}{J2109.4+1437$^{\dagger}$} & \multicolumn{1}{c|}{63.689} & \multicolumn{1}{c|}{-21.881} & \multicolumn{1}{c|}{-2.28} & \multicolumn{1}{c|}{1.00} & \multicolumn{1}{c|}{-0.70} & \multicolumn{1}{c|}{-0.08} & \multicolumn{1}{c|}{-0.62} & \multicolumn{1}{c|}{1.00} & \multicolumn{1}{c|}{0.8806} & \multicolumn{1}{c|}{0.8886} & \multicolumn{1}{c|}{fsrq} \\ 
\hline 
\multicolumn{1}{|l||}{J2250.3+1747$^{\dagger}$} & \multicolumn{1}{c|}{86.354} & \multicolumn{1}{c|}{-36.331} & \multicolumn{1}{c|}{-1.02} & \multicolumn{1}{c|}{1.08} & \multicolumn{1}{c|}{-0.28} & \multicolumn{1}{c|}{-0.90} & \multicolumn{1}{c|}{0.62} & \multicolumn{1}{c|}{1.00} & \multicolumn{1}{c|}{0.8135} & \multicolumn{1}{c|}{1.2062} & \multicolumn{1}{c|}{fsrq} \\ 
\hline

  \end{tabular}
  \label{tab:FermiAnalysisTable}
\end{table*}
\normalsize

\begin{table}
  \centering
  \caption[]{Summary of results obtained when applying the classifiers to
    the BCUs and to the blazar candidates from case A.
    $N_\text{s}$ is the number of sources in the different target samples.
    $N^{\text{bll}}_\text{c}$, $N^{\text{bll}}_\text{false}$ and 
    $N^{\text{bll}}_\text{miss}$ are respectively the number of \bll candidates, 
    the corresponding estimated number of false associations and
    the estimated number of \blls missed by the classifiers.
    Similarly, 
    $N^{\text{fsrq}}_\text{c}$, $N^{\text{fsrq}}_\text{false}$
    and $N^{\text{fsrq}}_\text{miss}$ are 
    the number of \fsrq candidates, the corresponding estimated number of
    false associations and
    the estimated number of \fsrqs missed by the classifiers, respectively.
    $N_\text{false}$  and $N_\text{miss}$ are obtained as explained in Table~\ref{tab:FermiAnalysisResults}.}
  \label{tab:FermiBllVsFsrqAnalysisPerf}
  \renewcommand{\tabcolsep}{1pt}
  \begin{tabular}{|c|l|c|c|c|c|c|c|c|}

    \hline

    \multirow{1}{*}{Target}
    & Flag
    & $N_\text{s}$
    & $N^{\text{bll}}_\text{c}$ 
    & $N^{\text{bll}}_\text{false}$
    & $N^{\text{bll}}_\text{miss}$ 
    & $N^{\text{fsrq}}_\text{c}$ 
    & $N^{\text{fsrq}}_\text{false}$ 
    & $N^{\text{fsrq}}_\text{miss}$ \\ \cline{1-9}

    \multirow{1}{*}{BCUs}
    & no
    & 486
    & 295
    & \SI{13.3}{}
    & \SI{54.2}{}
    & 146
    & \SI{39.3}{}
    & \SI{19.7}{} \\ \cline{1-9}

    \multirow{1}{*}{Blazar candidates}
    & no
    & 417
    & 214
    & \SI{16.1}{}
    & \SI{38.1}{}
    & 149
    & \SI{30.2}{}
    & \SI{21.9}{} \\ \cline{1-9}

  \end{tabular}

\end{table}

For case B, the classifiers were applied to 903 blazar candidates (only sources with no flag were considered),
486 being labelled as BCU in the 3FGL catalogue and 417 being labelled as blazar candidates in our case A study.
From this we obtained a list of 509 \bll candidates with an estimated number of $\sim$29 false associations and a list of 295 \fsrq
candidates with an estimated number of $\sim$70 false associations, hence leaving 99 blazars with uncertain type\footnote{By definition, the type of a source 
is considered as uncertain if $\zeta_{\text{BDT}}^{\star\text{,fsrq}} < \zeta_{\text{BDT}}< \zeta_{\text{BDT}}^{\star\text{,bll}}$
or $\zeta_{\text{MLP}}^{\star\text{,fsrq}} < \zeta_{\text{MLP}}< \zeta_{\text{MLP}}^{\star\text{,bll}}$.}.
Details are given in Table~\ref{tab:FermiBllVsFsrqAnalysisPerf} and a short sample of sources is shown on Table~\ref{tab:FermiBcuAnalysisTable}.

\scriptsize
\renewcommand{\tabcolsep}{2pt}
\begin{table*}
  \caption{Results of the \bll/\fsrq classifications (case B) for the first ten
    BCUs with no flag. 
    The columns correspond respectively to the 3FGL source name,
    the galactic coordinates of the source ($l$,$b$),
    the values of the five discriminant parameters, the values of the output parameters $\zeta$
    built with the BDT and MLP classifiers and the assignation for the \bll or \fsrq nature of the source (bll for \bll, fsrq for \fsrq and unc for uncertain).
    This table is available in its entirety on
    \protect\href{https://unidgamma.in2p3.fr}{https://unidgamma.in2p3.fr}.}

  \centering

  \begin{tabular}{|l||c|c|c|c|c|c|c|c|c|c|c|}

    \hline

    \multicolumn{1}{|l||}{3FGL name}
    & \multicolumn{1}{c|}{$l$ (\SI{}{\degree})}
    & \multicolumn{1}{c|}{$b$ (\SI{}{\degree})}
    & \multicolumn{1}{c|}{$\gamma$}
    & \multicolumn{1}{c|}{$\log_{10} E_p$}
    & \multicolumn{1}{c|}{$\log_{10} \widetilde{\text{TS}}$}
    & \multicolumn{1}{c|}{$\text{HR}_{23}$}
    & \multicolumn{1}{c|}{$\text{HR}_{34}$}
    & \multicolumn{1}{c|}{$\zeta_{\text{BDT}}$}
    & \multicolumn{1}{c|}{$\zeta_{\text{MLP}}$} 
    & \multicolumn{1}{c|}{type} \\
    \hline
    \hline

    \multicolumn{1}{|l||}{J0002.2-4152} & \multicolumn{1}{c|}{334.070} & \multicolumn{1}{c|}{-72.143} & \multicolumn{1}{c|}{2.09} & \multicolumn{1}{c|}{3.29} & \multicolumn{1}{c|}{1.04} & \multicolumn{1}{c|}{-0.35} & \multicolumn{1}{c|}{0.56} & \multicolumn{1}{c|}{0.0576} & \multicolumn{1}{c|}{0.6513} & \multicolumn{1}{c|}{unc} \\ 
\hline 
\multicolumn{1}{|l||}{J0003.2-5246} & \multicolumn{1}{c|}{318.976} & \multicolumn{1}{c|}{-62.825} & \multicolumn{1}{c|}{1.90} & \multicolumn{1}{c|}{3.57} & \multicolumn{1}{c|}{0.90} & \multicolumn{1}{c|}{1.00} & \multicolumn{1}{c|}{0.15} & \multicolumn{1}{c|}{0.3141} & \multicolumn{1}{c|}{1.1037} & \multicolumn{1}{c|}{bll} \\ 
\hline 
\multicolumn{1}{|l||}{J0003.8-1151} & \multicolumn{1}{c|}{84.432} & \multicolumn{1}{c|}{-71.084} & \multicolumn{1}{c|}{2.02} & \multicolumn{1}{c|}{3.36} & \multicolumn{1}{c|}{1.06} & \multicolumn{1}{c|}{0.17} & \multicolumn{1}{c|}{-0.23} & \multicolumn{1}{c|}{0.0951} & \multicolumn{1}{c|}{0.8209} & \multicolumn{1}{c|}{bll} \\ 
\hline 
\multicolumn{1}{|l||}{J0014.6+6119} & \multicolumn{1}{c|}{118.530} & \multicolumn{1}{c|}{-1.239} & \multicolumn{1}{c|}{1.89} & \multicolumn{1}{c|}{3.71} & \multicolumn{1}{c|}{0.84} & \multicolumn{1}{c|}{0.46} & \multicolumn{1}{c|}{0.28} & \multicolumn{1}{c|}{0.4634} & \multicolumn{1}{c|}{0.8286} & \multicolumn{1}{c|}{bll} \\ 
\hline 
\multicolumn{1}{|l||}{J0015.7+5552} & \multicolumn{1}{c|}{117.909} & \multicolumn{1}{c|}{-6.649} & \multicolumn{1}{c|}{2.11} & \multicolumn{1}{c|}{3.48} & \multicolumn{1}{c|}{0.77} & \multicolumn{1}{c|}{-0.29} & \multicolumn{1}{c|}{-0.10} & \multicolumn{1}{c|}{0.4029} & \multicolumn{1}{c|}{1.0609} & \multicolumn{1}{c|}{bll} \\ 
\hline 
\multicolumn{1}{|l||}{J0017.2-0643} & \multicolumn{1}{c|}{99.669} & \multicolumn{1}{c|}{-68.036} & \multicolumn{1}{c|}{2.12} & \multicolumn{1}{c|}{3.28} & \multicolumn{1}{c|}{0.78} & \multicolumn{1}{c|}{0.39} & \multicolumn{1}{c|}{-0.27} & \multicolumn{1}{c|}{0.5943} & \multicolumn{1}{c|}{1.0111} & \multicolumn{1}{c|}{bll} \\ 
\hline 
\multicolumn{1}{|l||}{J0019.1-5645} & \multicolumn{1}{c|}{311.732} & \multicolumn{1}{c|}{-59.822} & \multicolumn{1}{c|}{2.39} & \multicolumn{1}{c|}{3.04} & \multicolumn{1}{c|}{0.97} & \multicolumn{1}{c|}{0.06} & \multicolumn{1}{c|}{-0.12} & \multicolumn{1}{c|}{-0.2171} & \multicolumn{1}{c|}{0.3723} & \multicolumn{1}{c|}{fsrq} \\ 
\hline 
\multicolumn{1}{|l||}{J0028.6+7507} & \multicolumn{1}{c|}{121.437} & \multicolumn{1}{c|}{12.313} & \multicolumn{1}{c|}{2.34} & \multicolumn{1}{c|}{3.07} & \multicolumn{1}{c|}{0.71} & \multicolumn{1}{c|}{-0.14} & \multicolumn{1}{c|}{0.10} & \multicolumn{1}{c|}{0.3046} & \multicolumn{1}{c|}{0.6484} & \multicolumn{1}{c|}{bll} \\ 
\hline 
\multicolumn{1}{|l||}{J0030.2-1646} & \multicolumn{1}{c|}{96.521} & \multicolumn{1}{c|}{-78.546} & \multicolumn{1}{c|}{1.65} & \multicolumn{1}{c|}{3.70} & \multicolumn{1}{c|}{0.85} & \multicolumn{1}{c|}{1.00} & \multicolumn{1}{c|}{0.19} & \multicolumn{1}{c|}{0.4917} & \multicolumn{1}{c|}{0.9672} & \multicolumn{1}{c|}{bll} \\ 
\hline 
\multicolumn{1}{|l||}{J0030.7-0209} & \multicolumn{1}{c|}{110.859} & \multicolumn{1}{c|}{-64.544} & \multicolumn{1}{c|}{2.38} & \multicolumn{1}{c|}{2.84} & \multicolumn{1}{c|}{1.37} & \multicolumn{1}{c|}{-0.24} & \multicolumn{1}{c|}{-0.29} & \multicolumn{1}{c|}{-0.4396} & \multicolumn{1}{c|}{-0.0341} & \multicolumn{1}{c|}{fsrq} \\ 
\hline

  \end{tabular}

  \label{tab:FermiBcuAnalysisTable}
\end{table*}
\normalsize

The lists of blazar candidates obtained in this work are available at \protect\href{https://unidgamma.in2p3.fr}{https://unidgamma.in2p3.fr}
in FITS format.

\section{Discussion and conclusions}
\label{Sec:Discussion}

The work presented here is in the continuity of previous studies
\citep{Ackermann:2012ab, Mirabal2012, Doert:2013aa, Hassan:2013aa} which used machine-learning algorithms
based on parameters from different Fermi/LAT catalogues (1FGL, 2FGL) to address
the question of the nature (blazar or other) of unassociated sources
or the nature (\bll or \fsrq) of blazars whose type is undetermined\footnote{
  It is not straightforward to compare the list of candidates provided by studies based on different
  Fermi/LAT catalogues (1FGL, 2FGL, 3FGL),
  the number of unassociated sources and the number of blazars with
  undetermined type being significantly different from one catalogue to another.
  To keep track of previous works we will indicate in
  Table \ref{tab:FermiAnalysisTable} those of our candidates that have been proposed elsewhere.}.
The specificity of this work, beyond the fact that it deals with the recently
published 3FGL catalogue, is that it shows how performance of classifiers differ for flagged or non-flagged sources,
and it provides for each list of candidates an estimation of the number of false associations.

This study provides a list of 497 blazar candidates, with an expected number of false associations $\sim$18
  (not including the 98 low galactic latitude flagged candidates, for which we expect a high number of false associations).
  This represents a substantial contribution to the knowledge of the $\gamma$-ray emitting blazars population,
  and complements the population of 1559 blazars in the 3LAC catalogue.  

Similarly to our case A study, \citet{Saz-Parkinson:2016aa} in a recently published paper tackle
the question of the nature of the 3FGL unassociated sources.
Their work is based on a combination of a random forest 
and a logistic regression method, trained using a set of nine
discriminant parameters to separate samples of well identified blazars and
pulsars. 
Among their selected parameters five have a corresponding parameter in our study with the 
same physical content, but in our case two were corrected to reduce the flux dependency of their separation
power ($\widetilde{\sigma}_c$ and  $\widetilde{\text{TS}}$, following a prescription of \citet{Doert:2013aa}); 
their remaining parameters ($\text{HR}_{12}$ and $\text{HR}_{45}$) were
discarded in our work as it is likely that they introduce biases in the performance of classifiers,
specially when applied to low flux sources (see  Section~\ref{Sec:DiscriParams}).
In addition, we note that the effect on the classifier behaviour of flagged sources (present in their training and test sample
or in the sample of unassociated sources) was not taken into account.
Also, the sample of sources labelled as SNR or PWN was not used for their classifier training while this kind of sources
 represent a non-negligible fraction of galactic sources.
 Applied to the set of unassociated sources, their classifiers give a list of 559 blazar candidates,
 with no indication of the expected number of false associations\footnote{The performance corresponding to the combination of
  their two classifiers decisions is not provided, while it is necessary for an estimation of the number of false associations.}.
Setting aside the sample of low galactic latitude sources (dominated by flagged sources,
potentially including SNR and PWN), they have 481 sources with galactic latitude $|b| > \SI{10}{\degree}$,
444 ($\sim$90\%) being also in our corresponding sample of 497 candidates. 
We note however that the difference ($\sim$10\%) is much higher than our expected number of false associations, which is $\sim$18.

Concerning the list of \bll and \fsrq candidates resulting from this work (case B),
we note a clear dominance of \blls, which represent $\sim$$\SI{63}{\percent}$.
This is close to the \bll dominance already observed in the 3LAC catalogue ($\sim$$\SI{59}{\percent}$)\footnote{
  We note also
  the continuously increasing fraction of \blls among the blazars
  of known type in the EGRET/Fermi-LAT energy range.
  From $\sim$$\SI{25}{\percent}$ in the Third EGRET Catalog \citep{Hartman:1999aa},
  it has increased to $\sim$$\SI{50}{\percent}$,
  $\sim$$\SI{56}{\percent}$ and $\sim$$\SI{59}{\percent}$,
  respectively in the 1LAC \citep{Abdo:2010ac}, 2LAC \citep{Ackermann:2011aa}
  and 3LAC \citep{Ackermann:2015aa} catalogues.
  Such an evolution is probably the reflect of the sensitivity improvement,
  specially in the GeV domain, first when passing from EGRET to Fermi-LAT
  and second due to the evolution of the analyses methods used in
  the different Fermi-LAT AGN catalogues.}.
Putting together the lists of  \blls and \fsrqs from 3LAC catalogue and from this work,
we obtain a sample of 1113 \blls and 709 \fsrqs. \blls represent then $\sim$$\SI{61}{\percent}$.

In addition, an interesting comparison can be made between the population of \blls and \fsrqs of the
3LAC catalogue and the population of our \bll and \fsrq 
candidates in terms of the position of their synchrotron peak frequencies.
For that, we considered only our candidates which were initially labelled as BCUs because only 
those have available information about the synchrotron peak frequencies in the 3FGL catalogue. 
Using the HSP, ISP and LSP definitions of \citet{Ackermann:2015aa}, corresponding respectively to 
high-synchrotron-peaked, intermediate-synchrotron-peaked and low-synchrotron-peaked,
we note that our population of \bll candidates is dominated by HSP (46\% HSP)
as it is the case for the \blls in the 3LAC catalogue (43\% HSP).
Similarly, our population of \fsrq candidates 
and the \fsrqs in the 3LAC catalogue are both dominated by
LSP (78\% and 88\%, respectively).

The lists of \bll and \fsrq candidates resulting from this work can be compared to those recently obtained by \citet{Chiaro:2016aa}.
Using a single classifier (MLP) built only from variability features to separate \blls and \fsrqs,
they obtained interesting performance for \blls,
with true and false positive rates of $\sim$$\SI{84}{\percent}$ and $\sim$$\SI{5}{\percent}$ (compared to $\sim$$\SI{84}{\percent}$
and $\sim$$\SI{9}{\percent}$ in our case B study). For \fsrqs they obtained true and false positive rates of $\sim$$\SI{69}{\percent}$
and $\sim$$\SI{12}{\percent}$ ($\sim$$\SI{84}{\percent}$ and $\sim$$\SI{11}{\percent}$ in our study).
Applied to the BCUs in the 3FGL catalogue, their classifier provides a list of 314 \bll and 113 \fsrq candidates.
The comparison with our corresponding 295 \bll candidates shows a good agreement,
as $\sim$$\SI{91}{\percent}$ of our candidates are seen also as \bll by
\citet{Chiaro:2016aa},
$\sim$$\SI{6}{\percent}$ are still undetermined and only $\sim$$\SI{3}{\percent}$ obtain an \fsrq label.
This $\sim$$\SI{3}{\percent}$ represent approximately nine sources, which is close to our expected number of false associations, $\sim$13. 
A poorer agreement is found for the \fsrq candidates.
Among our 146 \fsrq candidates, only 92 are seen as \fsrqs by \citet{Chiaro:2016aa}, while 23 are seen as \blls and 31 remain of undetermined type.
Interestingly, considering the distribution of the normalised variability $\widetilde{\text{TS}}$, which carries in our case B study the information on temporal variability,
the 23 sources for which we don't find agreement with \citet{Chiaro:2016aa} are located in a region corresponding to the overlap between \blls and \fsrqs.
However, considering different combinations of our selected spectral parameters, these 23 sources appear clearly as being preferentially \fsrqs than \blls.
This illustrates the interest of taking into account spectral parameters for \bll/\fsrq separation purposes.

Finally, an interesting validation of the quality of our results is provided by a recent campaign 
of spectroscopic observations performed by \cite{Alvarez-Crespo:2016aa} and \cite{Alvarez-Crespo:2016ab}. 
They measured with different telescopes the optical spectra of 60 $\gamma$-ray blazar candidates selected on 
the basis of their IR colours or their low radio frequency spectra and belonging to different Fermi/LAT catalogues 
(principally BCUs or potential counterparts for unassociated sources). 
Their list contains five unassociated sources and 26 BCUs with no flag in the 3FGL catalogue.
Our case B study found a high-confidence classification for 27 out of these 31 sources as BL Lacs or FSRQs,
25 of which are spectroscopically confirmed by \cite{Alvarez-Crespo:2016aa} and \cite{Alvarez-Crespo:2016ab}. 
We note that one of the two remaining sources is WISE~J014935.28+860115.4, which shows an optical spectrum dominated by the host galaxy
and is identified as \bll/galaxy by \cite{Alvarez-Crespo:2016aa}. The other is WISEA~J122127.20-062847.8, and is not
clearly established as the correct counterpart of the $\gamma$-ray source 3FGL~J1221.5–0632 \citep{Alvarez-Crespo:2016ab,Massaro:2013ab}.

This study contributes significantly to increase and better constrain the sample of $\gamma$-ray blazars, 
based on the $\gamma$-ray detections performed by \fermi/\lat in four years of observation.
We expect that it will trigger multiwavelength follow-ups to assert the veracity of the proposed associations.
Additionally, the blazar candidate samples
might be of particular interest for contemporary very high energy $\gamma$-ray experiments using 
the imaging atmospheric Cherenkov technique such as H.E.S.S., MAGIC, and VERITAS, and later for
the next generation of arrays currently under construction by the Cherenkov Telescope Array (CTA) Consortium.
At present, population studies of very high energy blazars are indeed
limited by the small number of detected sources ($\sim$70), which is strongly dominated by \bll objects.

\begin{acknowledgements}
  We thank Catherine Boisson (LUTH, Observatoire de Paris) and Arache
  Djannati-Ata\"i (APC, IN2P3/CNRS) 
  for useful discussions at different levels of this work.
  We also thank the referee for useful comments
  and suggestions that led to improvements in the manuscript.
  We acknowledge the financial support of the APC laboratory.
  This study used TMVA\footnote{\protect\href{http://tmva.sourceforge.net/}{http://tmva.sourceforge.net/}} \citep{Hoecker:2007aa}: an open-source toolkit
  for multivariate data analysis.
  STILTS\footnote{\href{http://www.starlink.ac.uk/stilts/}{http://www.starlink.ac.uk/stilts/}} \citep{Taylor:2006aa}
  was used to manipulate tabular data and
  to cross match catalogues.
\end{acknowledgements}

\bibliographystyle{aa} 
\bibliography{biblio} 

\begin{thebibliography}{33}
\expandafter\ifx\csname natexlab\endcsname\relax\def\natexlab#1{#1}\fi

\bibitem[{{Abdo} {et~al.}(2010{\natexlab{a}}){Abdo}, {Ackermann}, {Ajello},
  {Allafort}, {Antolini}, {Atwood}, {Axelsson}, {Baldini}, {Ballet},
  {Barbiellini}, {Bastieri}, {Baughman}, {Bechtol}, {Bellazzini}, {Berenji},
  {Blandford}, {Bloom}, {Bogart}, {Bonamente}, {Borgland}, {Bouvier},
  {Bregeon}, {Brez}, {Brigida}, {Bruel}, {Buehler}, {Burnett}, {Buson},
  {Caliandro}, {Cameron}, {Cannon}, {Caraveo}, {Carrigan}, {Casandjian},
  {Cavazzuti}, {Cecchi}, {{\c C}elik}, {Celotti}, {Charles}, {Chekhtman},
  {Chen}, {Cheung}, {Chiang}, {Ciprini}, {Claus}, {Cohen-Tanugi}, {Conrad},
  {Costamante}, {Cotter}, {Cutini}, {D'Elia}, {Dermer}, {de Angelis}, {de
  Palma}, {De Rosa}, {Digel}, {Silva}, {Drell}, {Dubois}, {Dumora}, {Escande},
  {Farnier}, {Favuzzi}, {Fegan}, {Ferrara}, {Focke}, {Fortin}, {Frailis},
  {Fukazawa}, {Funk}, {Fusco}, {Gargano}, {Gasparrini}, {Gehrels}, {Germani},
  {Giebels}, {Giglietto}, {Giommi}, {Giordano}, {Giroletti}, {Glanzman},
  {Godfrey}, {Grandi}, {Grenier}, {Grondin}, {Grove}, {Guiriec}, {Hadasch},
  {Harding}, {Hayashida}, {Hays}, {Healey}, {Hill}, {Horan}, {Hughes},
  {Iafrate}, {Itoh}, {J{\'o}hannesson}, {Johnson}, {Johnson}, {Johnson},
  {Johnson}, {Kamae}, {Katagiri}, {Kataoka}, {Kawai}, {Kerr}, {Kn{\"o}dlseder},
  {Kuss}, {Lande}, {Latronico}, {Lavalley}, {Lemoine-Goumard}, {Llena Garde},
  {Longo}, {Loparco}, {Lott}, {Lovellette}, {Lubrano}, {Madejski}, {Makeev},
  {Malaguti}, {Massaro}, {Mazziotta}, {McConville}, {McEnery}, {McGlynn},
  {Michelson}, {Mitthumsiri}, {Mizuno}, {Moiseev}, {Monte}, {Monzani},
  {Morselli}, {Moskalenko}, {Murgia}, {Nolan}, {Norris}, {Nuss}, {Ohno},
  {Ohsugi}, {Omodei}, {Orlando}, {Ormes}, {Ozaki}, {Paneque}, {Panetta},
  {Parent}, {Pelassa}, {Pepe}, {Pesce-Rollins}, {Piranomonte}, {Piron},
  {Porter}, {Rain{\`o}}, {Rando}, {Razzano}, {Reimer}, {Reimer}, {Reposeur},
  {Ripken}, {Ritz}, {Rodriguez}, {Romani}, {Roth}, {Ryde}, {Sadrozinski},
  {Sanchez}, {Sander}, {Saz Parkinson}, {Scargle}, {Sgr{\`o}}, {Shaw},
  {Siskind}, {Smith}, {Spandre}, {Spinelli}, {Starck}, {Stawarz}, {Strickman},
  {Suson}, {Tajima}, {Takahashi}, {Takahashi}, {Tanaka}, {Taylor}, {Thayer},
  {Thayer}, {Thompson}, {Tibaldo}, {Torres}, {Tosti}, {Tramacere}, {Ubertini},
  {Uchiyama}, {Usher}, {Vasileiou}, {Vilchez}, {Villata}, {Vitale}, {Waite},
  {Wallace}, {Wang}, {Winer}, {Wood}, {Yang}, {Ylinen}, \&
  {Ziegler}}]{Abdo:2010ac}
{Abdo}, A.~A., {Ackermann}, M., {Ajello}, M., {et~al.} 2010{\natexlab{a}},
  \apj, 715, 429

\bibitem[{{Abdo} {et~al.}(2010{\natexlab{b}}){Abdo}, {Ackermann}, {Ajello},
  {Allafort}, {Antolini}, {Atwood}, {Axelsson}, {Baldini}, {Ballet},
  {Barbiellini}, \& et~al.}]{Abdo:2010ab}
{Abdo}, A.~A., {Ackermann}, M., {Ajello}, M., {et~al.} 2010{\natexlab{b}},
  \apjs, 188, 405

\bibitem[{{Acero} {et~al.}(2015){Acero}, {Ackermann}, {Ajello}, {Albert},
  {Atwood}, {Axelsson}, {Baldini}, {Ballet}, {Barbiellini}, {Bastieri},
  {Belfiore}, {Bellazzini}, {Bissaldi}, {Blandford}, {Bloom}, {Bogart},
  {Bonino}, {Bottacini}, {Bregeon}, {Britto}, {Bruel}, {Buehler}, {Burnett},
  {Buson}, {Caliandro}, {Cameron}, {Caputo}, {Caragiulo}, {Caraveo},
  {Casandjian}, {Cavazzuti}, {Charles}, {Chaves}, {Chekhtman}, {Cheung},
  {Chiang}, {Chiaro}, {Ciprini}, {Claus}, {Cohen-Tanugi}, {Cominsky}, {Conrad},
  {Cutini}, {D'Ammando}, {de Angelis}, {DeKlotz}, {de Palma}, {Desiante},
  {Digel}, {Di Venere}, {Drell}, {Dubois}, {Dumora}, {Favuzzi}, {Fegan},
  {Ferrara}, {Finke}, {Franckowiak}, {Fukazawa}, {Funk}, {Fusco}, {Gargano},
  {Gasparrini}, {Giebels}, {Giglietto}, {Giommi}, {Giordano}, {Giroletti},
  {Glanzman}, {Godfrey}, {Grenier}, {Grondin}, {Grove}, {Guillemot}, {Guiriec},
  {Hadasch}, {Harding}, {Hays}, {Hewitt}, {Hill}, {Horan}, {Iafrate}, {Jogler},
  {J{\'o}hannesson}, {Johnson}, {Johnson}, {Johnson}, {Johnson}, {Kamae},
  {Kataoka}, {Katsuta}, {Kuss}, {La Mura}, {Landriu}, {Larsson}, {Latronico},
  {Lemoine-Goumard}, {Li}, {Li}, {Longo}, {Loparco}, {Lott}, {Lovellette},
  {Lubrano}, {Madejski}, {Massaro}, {Mayer}, {Mazziotta}, {McEnery},
  {Michelson}, {Mirabal}, {Mizuno}, {Moiseev}, {Mongelli}, {Monzani},
  {Morselli}, {Moskalenko}, {Murgia}, {Nuss}, {Ohno}, {Ohsugi}, {Omodei},
  {Orienti}, {Orlando}, {Ormes}, {Paneque}, {Panetta}, {Perkins},
  {Pesce-Rollins}, {Piron}, {Pivato}, {Porter}, {Racusin}, {Rando}, {Razzano},
  {Razzaque}, {Reimer}, {Reimer}, {Reposeur}, {Rochester}, {Romani},
  {Salvetti}, {S{\'a}nchez-Conde}, {Saz Parkinson}, {Schulz}, {Siskind},
  {Smith}, {Spada}, {Spandre}, {Spinelli}, {Stephens}, {Strong}, {Suson},
  {Takahashi}, {Takahashi}, {Tanaka}, {Thayer}, {Thayer}, {Thompson},
  {Tibaldo}, {Tibolla}, {Torres}, {Torresi}, {Tosti}, {Troja}, {Van Klaveren},
  {Vianello}, {Winer}, {Wood}, {Wood}, {Zimmer}, \& {Fermi-LAT
  Collaboration}}]{Acero:2015aa}
{Acero}, F., {Ackermann}, M., {Ajello}, M., {et~al.} 2015, \apjs, 218, 23

\bibitem[{{Acero} {et~al.}(2013){Acero}, {Donato}, {Ojha}, {Stevens},
  {Edwards}, {Ferrara}, {Blanchard}, {Lovell}, \& {Thompson}}]{Acero:2013aa}
{Acero}, F., {Donato}, D., {Ojha}, R., {et~al.} 2013, \apj, 779, 133

\bibitem[{{Ackermann} {et~al.}(2011){Ackermann}, {Ajello}, {Allafort},
  {Antolini}, {Atwood}, {Axelsson}, {Baldini}, {Ballet}, {Barbiellini},
  {Bastieri}, {Bechtol}, {Bellazzini}, {Berenji}, {Blandford}, {Bloom},
  {Bonamente}, {Borgland}, {Bottacini}, {Bouvier}, {Bregeon}, {Brigida},
  {Bruel}, {Buehler}, {Burnett}, {Buson}, {Caliandro}, {Cameron}, {Caraveo},
  {Casandjian}, {Cavazzuti}, {Cecchi}, {Charles}, {Cheung}, {Chiang},
  {Ciprini}, {Claus}, {Cohen-Tanugi}, {Conrad}, {Costamante}, {Cutini}, {de
  Angelis}, {de Palma}, {Dermer}, {Digel}, {Silva}, {Drell}, {Dubois},
  {Escande}, {Favuzzi}, {Fegan}, {Ferrara}, {Finke}, {Focke}, {Fortin},
  {Frailis}, {Fukazawa}, {Funk}, {Fusco}, {Gargano}, {Gasparrini}, {Gehrels},
  {Germani}, {Giebels}, {Giglietto}, {Giommi}, {Giordano}, {Giroletti},
  {Glanzman}, {Godfrey}, {Grenier}, {Grove}, {Guiriec}, {Gustafsson},
  {Hadasch}, {Hayashida}, {Hays}, {Healey}, {Horan}, {Hou}, {Hughes},
  {Iafrate}, {J{\'o}hannesson}, {Johnson}, {Johnson}, {Kamae}, {Katagiri},
  {Kataoka}, {Kn{\"o}dlseder}, {Kuss}, {Lande}, {Larsson}, {Latronico},
  {Longo}, {Loparco}, {Lott}, {Lovellette}, {Lubrano}, {Madejski}, {Mazziotta},
  {McConville}, {McEnery}, {Michelson}, {Mitthumsiri}, {Mizuno}, {Moiseev},
  {Monte}, {Monzani}, {Moretti}, {Morselli}, {Moskalenko}, {Murgia},
  {Nakamori}, {Naumann-Godo}, {Nolan}, {Norris}, {Nuss}, {Ohno}, {Ohsugi},
  {Okumura}, {Omodei}, {Orienti}, {Orlando}, {Ormes}, {Ozaki}, {Paneque},
  {Parent}, {Pesce-Rollins}, {Pierbattista}, {Piranomonte}, {Piron}, {Pivato},
  {Porter}, {Rain{\`o}}, {Rando}, {Razzano}, {Razzaque}, {Reimer}, {Reimer},
  {Ritz}, {Rochester}, {Romani}, {Roth}, {Sanchez}, {Sbarra}, {Scargle},
  {Schalk}, {Sgr{\`o}}, {Shaw}, {Siskind}, {Spandre}, {Spinelli}, {Strong},
  {Suson}, {Tajima}, {Takahashi}, {Takahashi}, {Tanaka}, {Thayer}, {Thayer},
  {Thompson}, {Tibaldo}, {Tinivella}, {Torres}, {Tosti}, {Troja}, {Uchiyama},
  {Vandenbroucke}, {Vasileiou}, {Vianello}, {Vitale}, {Waite}, {Wallace},
  {Wang}, {Winer}, {Wood}, {Wood}, \& {Zimmer}}]{Ackermann:2011aa}
{Ackermann}, M., {Ajello}, M., {Allafort}, A., {et~al.} 2011, \apj, 743, 171

\bibitem[{{Ackermann} {et~al.}(2012){Ackermann}, {Ajello}, {Allafort},
  {Antolini}, {Baldini}, {Ballet}, {Barbiellini}, \& et~al.}]{Ackermann:2012ab}
{Ackermann}, M., {Ajello}, M., {Allafort}, A., {et~al.} 2012, \apj, 753, 83

\bibitem[{{Ackermann} {et~al.}(2015){Ackermann}, {Ajello}, {Atwood}, {Baldini},
  {Ballet}, {Barbiellini}, {Bastieri}, {Becerra Gonzalez}, {Bellazzini},
  {Bissaldi}, {Blandford}, {Bloom}, {Bonino}, {Bottacini}, {Brandt}, {Bregeon},
  {Britto}, {Bruel}, {Buehler}, {Buson}, {Caliandro}, {Cameron}, {Caragiulo},
  {Caraveo}, {Carpenter}, {Casandjian}, {Cavazzuti}, {Cecchi}, {Charles},
  {Chekhtman}, {Cheung}, {Chiang}, {Chiaro}, {Ciprini}, {Claus},
  {Cohen-Tanugi}, {Cominsky}, {Conrad}, {Cutini}, {D'Abrusco}, {D'Ammando}, {de
  Angelis}, {Desiante}, {Digel}, {Di Venere}, {Drell}, {Favuzzi}, {Fegan},
  {Ferrara}, {Finke}, {Focke}, {Franckowiak}, {Fuhrmann}, {Fukazawa},
  {Furniss}, {Fusco}, {Gargano}, {Gasparrini}, {Giglietto}, {Giommi},
  {Giordano}, {Giroletti}, {Glanzman}, {Godfrey}, {Grenier}, {Grove},
  {Guiriec}, {Hewitt}, {Hill}, {Horan}, {Itoh}, {J{\'o}hannesson}, {Johnson},
  {Johnson}, {Kataoka}, {Kawano}, {Krauss}, {Kuss}, {La Mura}, {Larsson},
  {Latronico}, {Leto}, {Li}, {Li}, {Longo}, {Loparco}, {Lott}, {Lovellette},
  {Lubrano}, {Madejski}, {Mayer}, {Mazziotta}, {McEnery}, {Michelson},
  {Mizuno}, {Moiseev}, {Monzani}, {Morselli}, {Moskalenko}, {Murgia}, {Nuss},
  {Ohno}, {Ohsugi}, {Ojha}, {Omodei}, {Orienti}, {Orlando}, {Paggi}, {Paneque},
  {Perkins}, {Pesce-Rollins}, {Piron}, {Pivato}, {Porter}, {Rain{\`o}},
  {Rando}, {Razzano}, {Razzaque}, {Reimer}, {Reimer}, {Romani}, {Salvetti},
  {Schaal}, {Schinzel}, {Schulz}, {Sgr{\`o}}, {Siskind}, {Sokolovsky}, {Spada},
  {Spandre}, {Spinelli}, {Stawarz}, {Suson}, {Takahashi}, {Takahashi},
  {Tanaka}, {Thayer}, {Thayer}, {Tibaldo}, {Torres}, {Torresi}, {Tosti},
  {Troja}, {Uchiyama}, {Vianello}, {Winer}, {Wood}, \&
  {Zimmer}}]{Ackermann:2015aa}
{Ackermann}, M., {Ajello}, M., {Atwood}, W.~B., {et~al.} 2015, \apj, 810, 14

\bibitem[{{{\'A}lvarez Crespo} {et~al.}(2016{\natexlab{a}}){{\'A}lvarez
  Crespo}, {Masetti}, {Ricci}, {Landoni}, {Pati{\~n}o-{\'A}lvarez}, {Massaro},
  {D'Abrusco}, {Paggi}, {Chavushyan}, {Jim{\'e}nez-Bail{\'o}n}, {Torrealba},
  {Latronico}, {La Franca}, {Smith}, \& {Tosti}}]{Alvarez-Crespo:2016aa}
{{\'A}lvarez Crespo}, N., {Masetti}, N., {Ricci}, F., {et~al.}
  2016{\natexlab{a}}, \aj, 151, 32

\bibitem[{{{\'A}lvarez Crespo} {et~al.}(2016{\natexlab{b}}){{\'A}lvarez
  Crespo}, {Massaro}, {Milisavljevic}, {Landoni}, {Chavushyan},
  {Pati{\~n}o-{\'A}lvarez}, {Masetti}, {Jim{\'e}nez-Bail{\'o}n}, {Strader},
  {Chomiuk}, {Katagiri}, {Kagaya}, {Cheung}, {Paggi}, {D'Abrusco}, {Ricci}, {La
  Franca}, {Smith}, \& {Tosti}}]{Alvarez-Crespo:2016ab}
{{\'A}lvarez Crespo}, N., {Massaro}, F., {Milisavljevic}, D., {et~al.}
  2016{\natexlab{b}}, \aj, 151, 95

\bibitem[{{Brain} \& {Webb}(1999)}]{BrainWebb1999}
{Brain}, D. \& {Webb}, J.~I. 1999, in Proceedings of the 4th Australian
  Knowledge Acquisition Workshop, Sydney, NSW, pp. 117--128

\bibitem[{{Chiaro} {et~al.}(2016){Chiaro}, {Salvetti}, {La Mura}, {Giroletti},
  {Thompson}, \& {Bastieri}}]{Chiaro:2016aa}
{Chiaro}, G., {Salvetti}, D., {La Mura}, G., {et~al.} 2016, ArXiv e-prints
  [\eprint[arXiv]{1607.07822}]

\bibitem[{{D'Abrusco} {et~al.}(2014){D'Abrusco}, {Massaro}, {Paggi}, {Smith},
  {Masetti}, {Landoni}, \& {Tosti}}]{DAbrusco:2014aa}
{D'Abrusco}, R., {Massaro}, F., {Paggi}, A., {et~al.} 2014, \apjs, 215, 14

\bibitem[{{Doert} \& {Errando}(2014)}]{Doert:2013aa}
{Doert}, M. \& {Errando}, M. 2014, \apj, 782, 41

\bibitem[{Fawcett(2006)}]{Fawcett:2006}
Fawcett, T. 2006, Pattern Recogn. Lett., 27, 861

\bibitem[{{Ferrara} {et~al.}(2012){Ferrara}, {Ojha}, {Monzani}, \&
  {Omodei}}]{Ferrara:2012aa}
{Ferrara}, E.~C., {Ojha}, R., {Monzani}, M.~E., \& {Omodei}, N. 2012, ArXiv
  e-prints [\eprint[arXiv]{1206.2571}]

\bibitem[{{Freund} \& {Schapire}(1996)}]{Freund1996}
{Freund}, Y. \& {Schapire}, R.~E. 1996, in Machine Learning: Proceedings of the
  Thirteenth International Conference

\bibitem[{{Hartman} {et~al.}(1999){Hartman}, {Bertsch}, {Bloom}, {Chen},
  {Deines-Jones}, {Esposito}, {Fichtel}, {Friedlander}, {Hunter}, {McDonald},
  {Sreekumar}, {Thompson}, {Jones}, {Lin}, {Michelson}, {Nolan}, {Tompkins},
  {Kanbach}, {Mayer-Hasselwander}, {M{\"u}cke}, {Pohl}, {Reimer}, {Kniffen},
  {Schneid}, {von Montigny}, {Mukherjee}, \& {Dingus}}]{Hartman:1999aa}
{Hartman}, R.~C., {Bertsch}, D.~L., {Bloom}, S.~D., {et~al.} 1999, \apjs, 123,
  79

\bibitem[{{Hassan} {et~al.}(2013){Hassan}, {Mirabal}, {Contreras}, \&
  {Oya}}]{Hassan:2013aa}
{Hassan}, T., {Mirabal}, N., {Contreras}, J.~L., \& {Oya}, I. 2013, \mnras,
  428, 220

\bibitem[{{Hoecker} {et~al.}(2007){Hoecker}, {Speckmayer}, {Stelzer},
  {Therhaag}, {von Toerne}, {Voss}, {Backes}, {Carli}, {Cohen}, {Christov},
  {Dannheim}, {Danielowski}, {Henrot-Versille}, {Jachowski}, {Kraszewski},
  {Krasznahorkay}, {Kruk}, {Mahalalel}, {Ospanov}, {Prudent}, {Robert},
  {Schouten}, {Tegenfeldt}, {Voigt}, {Voss}, {Wolter}, \&
  {Zemla}}]{Hoecker:2007aa}
{Hoecker}, A., {Speckmayer}, P., {Stelzer}, J., {et~al.} 2007, ArXiv Physics
  e-prints [\eprint{physics/0703039}]

\bibitem[{{Massaro} {et~al.}(2011){Massaro}, {D'Abrusco}, {Ajello}, {Grindlay},
  \& {Smith}}]{Massaro:2011aa}
{Massaro}, F., {D'Abrusco}, R., {Ajello}, M., {Grindlay}, J.~E., \& {Smith},
  H.~A. 2011, \apjl, 740, L48

\bibitem[{{Massaro} {et~al.}(2013{\natexlab{a}}){Massaro}, {D'Abrusco},
  {Giroletti}, {Paggi}, {Masetti}, {Tosti}, {Nori}, \& {Funk}}]{Massaro:2013aa}
{Massaro}, F., {D'Abrusco}, R., {Giroletti}, M., {et~al.} 2013{\natexlab{a}},
  \apjs, 207, 4

\bibitem[{{Massaro} {et~al.}(2013{\natexlab{b}}){Massaro}, {D'Abrusco},
  {Paggi}, {Masetti}, {Giroletti}, {Tosti}, {Smith}, \&
  {Funk}}]{Massaro:2013ab}
{Massaro}, F., {D'Abrusco}, R., {Paggi}, A., {et~al.} 2013{\natexlab{b}},
  \apjs, 206, 13

\bibitem[{{Massaro} {et~al.}(2014){Massaro}, {D'Abrusco}, {Paggi}, {Masetti},
  {Giroletti}, {Tosti}, {Smith}, \& {Funk}}]{Massaro:2014aa}
{Massaro}, F., {D'Abrusco}, R., {Paggi}, A., {et~al.} 2014, ArXiv e-prints
  [\eprint[arXiv]{1404.5960}]

\bibitem[{{Massaro} {et~al.}(2012{\natexlab{a}}){Massaro}, {D'Abrusco},
  {Tosti}, {Ajello}, {Gasparrini}, {Grindlay}, \& {Smith}}]{Massaro:2012ac}
{Massaro}, F., {D'Abrusco}, R., {Tosti}, G., {et~al.} 2012{\natexlab{a}}, \apj,
  750, 138

\bibitem[{{Massaro} {et~al.}(2012{\natexlab{b}}){Massaro}, {D'Abrusco},
  {Tosti}, {Ajello}, {Paggi}, \& {Gasparrini}}]{Massaro:2012aa}
{Massaro}, F., {D'Abrusco}, R., {Tosti}, G., {et~al.} 2012{\natexlab{b}}, \apj,
  752, 61

\bibitem[{{Mirabal} {et~al.}(2012){Mirabal}, {Fr{\'{\i}}as-Martinez}, {Hassan},
  \& {Fr{\'{\i}}as-Martinez}}]{Mirabal2012}
{Mirabal}, N., {Fr{\'{\i}}as-Martinez}, V., {Hassan}, T., \&
  {Fr{\'{\i}}as-Martinez}, E. 2012, \mnras, 424

\bibitem[{{Nolan} {et~al.}(2012){Nolan}, {Abdo}, {Ackermann}, {Ajello},
  {Allafort}, {Antolini}, {Atwood}, {Axelsson}, {Baldini}, {Ballet}, \&
  et~al.}]{Nolan:2012aa}
{Nolan}, P.~L., {Abdo}, A.~A., {Ackermann}, M., {et~al.} 2012, \apjs, 199, 31

\bibitem[{{Paggi} {et~al.}(2014){Paggi}, {Massaro}, {D'Abrusco}, {Smith},
  {Masetti}, {Giroletti}, {Tosti}, \& {Funk}}]{Paggi:2014aa}
{Paggi}, A., {Massaro}, F., {D'Abrusco}, R., {et~al.} 2014, ArXiv e-prints
  [\eprint[arXiv]{1404.4631}]

\bibitem[{{Rumelhart} {et~al.}(1986){Rumelhart}, {Hinton}, \&
  {Williams}}]{Rumelhart1986}
{Rumelhart}, D.~E., {Hinton}, G.~E., \& {Williams}, R.~J. 1986, Nature

\bibitem[{{Saz Parkinson} {et~al.}(2016){Saz Parkinson}, {Xu}, {Yu},
  {Salvetti}, {Marelli}, \& {Falcone}}]{Saz-Parkinson:2016aa}
{Saz Parkinson}, P.~M., {Xu}, H., {Yu}, P.~L.~H., {et~al.} 2016, \apj, 820, 8

\bibitem[{{Sol} {et~al.}(2013){Sol}, {Zech}, {Boisson}, {Barres de Almeida},
  {Biteau}, {Contreras}, {Giebels}, {Hassan}, {Inoue}, {Katarzy{\'n}ski},
  {Krawczynski}, {Mirabal}, {Poutanen}, {Rieger}, {Totani}, {Benbow},
  {Cerruti}, {Errando}, {Fallon}, {de Gouveia Dal Pino}, {Hinton}, {Inoue},
  {Lenain}, {Neronov}, {Takahashi}, {Takami}, {White}, \& {CTA
  Consortium}}]{Sol:2013aa}
{Sol}, H., {Zech}, A., {Boisson}, C., {et~al.} 2013, Astroparticle Physics, 43,
  215

\bibitem[{{Taylor}(2006)}]{Taylor:2006aa}
{Taylor}, M.~B. 2006, in Astronomical Society of the Pacific Conference Series,
  Vol. 351, Astronomical Data Analysis Software and Systems XV, ed.
  C.~{Gabriel}, C.~{Arviset}, D.~{Ponz}, \& S.~{Enrique}, 666

\bibitem[{{Wright} {et~al.}(2010){Wright}, {Eisenhardt}, {Mainzer}, {Ressler},
  {Cutri}, {Jarrett}, {Kirkpatrick}, {Padgett}, {McMillan}, {Skrutskie},
  {Stanford}, {Cohen}, {Walker}, {Mather}, {Leisawitz}, {Gautier}, {McLean},
  {Benford}, {Lonsdale}, {Blain}, {Mendez}, {Irace}, {Duval}, {Liu}, {Royer},
  {Heinrichsen}, {Howard}, {Shannon}, {Kendall}, {Walsh}, {Larsen}, {Cardon},
  {Schick}, {Schwalm}, {Abid}, {Fabinsky}, {Naes}, \& {Tsai}}]{Wright:2010aa}
{Wright}, E.~L., {Eisenhardt}, P.~R.~M., {Mainzer}, A.~K., {et~al.} 2010, \aj,
  140, 1868

\end{thebibliography}

\end{document}